\documentclass[traditabstract]{aa}    

\newcommand{\hii}{H\,{\scriptsize II}}

\newcommand{\av}{A$_{\rm v}$}    
\newcommand{\dav}{$\Delta$A$_{\rm v}$}    
\usepackage{epsfig,graphicx,color}    
\usepackage{ctable}    
    
\usepackage{txfonts}    
\usepackage{ctable}    
\usepackage[toc,page]{appendix}    
    
\begin{document}

\title{Understanding star formation in molecular clouds}    
\subtitle{I. Effects of line-of-sight contamination on the column density structure }     
    
  \author{N. Schneider \inst{1,2}    
  \and V. Ossenkopf    \inst{3}    
  \and T. Csengeri     \inst{4}     
  \and R.S. Klessen    \inst{5,6,7}    
  \and C. Federrath    \inst{8,9}    
  \and P. Tremblin     \inst{10,11}  
  \and P. Girichidis   \inst{12}    
  \and S. Bontemps     \inst{1,2}    
  \and Ph. Andr\'e     \inst{13}               
  }      
 \institute{Univ. Bordeaux, LAB, UMR 5804, 33271 Floirac, France   
  \and   
  CNRS, LAB, UMR 5804, 33271 Floirac, France   
  \and   
  I.\,Physikalisches Institut, Universit\"at zu K\"oln, Z\"ulpicher Stra{\ss}e 77, 50937 K\"oln, Germany    
   \and  
 Max-Planck Institut f\"ur Radioastronomie, Auf dem H\"ugel 69, Bonn, Germany     
  \and   
   Universit\"at Heidelberg, Zentrum f\"ur Astronomie, Institut f\"ur Theoretische Astrophysik, 69120 Heidelberg, Germany    
  \and   
 Kavli Institute for Particle Astrophysics and Cosmology, Stanford University, SLAC National Accelerator Laboratory, CA 94025, USA 
  \and   
  Department of Astronomy and Astrophysics, University of California, Santa Cruz, CA 95064, USA  
  \and   
  Monash Centre for Astrophysics, School of Mathematical Sciences,  Monash University, VIC 3800, Australia    
  \and   
  Research School of Astronomy \& Astrophysics, The Australian National University, Canberra,  ACT 2611, Australia 
  \and   
  Astrophysics Group, University of Exeter, EX4 4QL Exeter, UK   
  \and   
  Maison de la Simulation, CEA-CNRS-INRIA-UPS-UVSQ, USR 3441, CEA Saclay, France 
  \and  
  Max-Planck Institut f\"ur Astrophysik, 85741 Garching, Germany   
  \and  
  IRFU/SAp CEA/DSM, Laboratoire AIM CNRS - Universit\'e Paris     
  Diderot, 91191 Gif-sur-Yvette, France    
}    
    
    
\mail{nicola.schneider@obs.u-bordeaux1.fr}    
    
\titlerunning{Understanding star formation - I. column density structure}    
\authorrunning{N. Schneider}    
    
    
\date{Received February 4, 2013; accepted December 15, 2014}    
    
\abstract {Column-density maps of molecular clouds are one of the most
  important observables in the context of molecular cloud- and
  star-formation (SF) studies.  With the {\sl Herschel} satellite it
  is now possible to precisely determine the column density from dust
  emission, which is the best tracer of the bulk of material in
  molecular clouds.  However, line-of-sight (LOS) contamination from
  fore- or background clouds can lead to overestimating the dust
  emission of molecular clouds, in particular for distant clouds.
  This implies values that are too high for column density and mass,
  which can potentially lead to an incorrect physical interpretation
  of the column density probability distribution function (PDF). In
  this paper, we use observations and simulations to demonstrate how
  LOS contamination affects the PDF.  We apply a first-order
  approximation (removing a constant level) to the molecular clouds of
  Auriga and Maddalena (low-mass star-forming), and Carina and NGC3603
  (both high-mass SF regions).  In perfect agreement with the
  simulations, we find that the PDFs become broader, the peak shifts
  to lower column densities, and the power-law tail of the PDF for
  higher column densities flattens after correction. All corrected
  PDFs have a lognormal part for low column densities with a peak at
  \av\, $\sim$2 mag, a deviation point (DP) from the lognormal at
  \av(DP)$\sim$4--5 mag, and a power-law tail for higher column
  densities. Assuming an equivalent spherical density distribution
  $\rho \propto r^{-\alpha}$, the slopes of the power-law tails
  correspond to $\alpha_{PDF}$ = 1.8, 1.75, and 2.5 for Auriga,
  Carina, and NGC3603.  These numbers agree within the uncertainties
  with the values of $\alpha \approx$ 1.5, 1.8, and 2.5 determined
  from the slope $\gamma$ (with $\alpha = 1 - \gamma$) obtained from
  the radial column density profiles ($N \propto r^\gamma$).  While
  \mbox{$\alpha \sim 1.5$--$2$} is consistent with a structure
  dominated by collapse (local free-fall collapse of individual cores
  and clumps and global collapse), the higher value of $\alpha >$ 2
  for NGC3603 requires a physical process that leads to additional
  compression (e.g., expanding ionization fronts).  From the small
  sample of our study, we find that clouds forming only low-mass stars
  and those also forming high-mass stars have slightly different
  values for their average column density (1.8 10$^{21}$cm$^{-2}$ vs.
  3.0 10$^{21}$cm$^{-2}$), and they display differences in the overall
  column density structure.  Massive clouds assemble more gas in
  smaller cloud volumes than low-mass SF ones.  However, for both
  cloud types, the transition of the PDF from lognormal shape into
  power-law tail is found at the same column density (at \av$\sim$4--5
  mag).  Low-mass and high-mass SF clouds then have the same low
  column density distribution, most likely dominated by supersonic
  turbulence. At higher column densities, collapse and external
  pressure can form the power-law tail. The relative importance of the
  two processes can vary between clouds and thus lead to the observed
  differences in PDF and column density structure.}
  
  \keywords{ISM: clouds -- ISM: structure -- ISM: dust, extinction -- Submillimeter: ISM -- Methods: data analysis    
          }    
    
   \maketitle    
    
      
\begin{table*}  
\caption{Molecular cloud parameters from {\sl Herschel} data. All values in parenthesis are the ones   
determined from the original maps before correction for LOS contamination. The last two lines give   
the average values (and standard deviation) from the corrected and uncorrected maps.} \label{table:summary}     
\begin{center}    
\begin{tabular}{lcccccccccl}    
  \hline \hline     
  & & & & & & & & & \\    
  Cloud & D & M &  $\Sigma$ & $\langle N(H_2) \rangle$ & $\Delta$A$_{\rm v}$ & A$_{\rm v,pk}$ & \av(DP)  & $\sigma_{\rm \eta}$ & $s$& $\alpha$  \\   
  & {\tiny [kpc]}  & {\tiny [10$^4$ M$_{\odot}$]} &   
  {\tiny [M$_\odot$ pc$^{-2}$]} & {\tiny [10$^{21}$ cm$^{-2}$]} & {\tiny [mag]}  &  {\tiny [mag]}   & {\tiny [mag]}  &  &  &   \\    
  & {\tiny (1)} & {\tiny (2)} & {\tiny (3)} & {\tiny (4)} & {\tiny (5)} & {\tiny (6)} & {\tiny (7)} & (8) & (9) & (10) \\    
 \hline         
  \multicolumn{3}{l}{ {\sl {\bf High-mass SF regions}}} & \\    
  {\tiny NGC3603}   & {\tiny 7.0}   & {\tiny 50.4 (97.1)} & {\tiny 60  (116)} &   
  {\tiny 3.24 (6.24)} & {\tiny 3.0$\pm$0.5} & {\tiny 1.6 (4.8)}  & {\tiny 4.9 (8.1)}  & {\tiny 0.52 (0.27)} & {\tiny -1.31 (-1.61)} &{\tiny 2.53 (2.24)} \\  
  {\tiny Carina}    & {\tiny 2.3}   & {\tiny 34.5 (59.5)} & {\tiny 50 (89)}   &   
  {\tiny 2.79 (4.79)} &  {\tiny 2.0$\pm$0.2} & {\tiny 2.6 (4.4)}  & {\tiny 5.5 (7.4)}  & {\tiny 0.38 (0.20)} & {\tiny -2.66 (-3.04)} &{\tiny 1.75 (1.66)} \\    
  \multicolumn{3}{l}{ {\sl {\bf Low-mass SF regions}}} & \\    
  {\tiny Maddalena} & {\tiny 2.2}   & {\tiny 35.2 (68.2)} & {\tiny 37 (76)}  &   
  {\tiny 2.13 (4.13)} & {\tiny  2.0$\pm$0.25} & {\tiny 1.9 (3.9)}      & {\tiny 4.9 (7.8)}& {\tiny 0.32 (0.20)} & {\tiny -3.69 (-5.21)} & {\tiny  1.54 (1.38)} \\    
  {\tiny Auriga}    & {\tiny 0.45}    & {\tiny 2.2 (3.7)}   & {\tiny 28 (47)}  &   
  {\tiny 1.51 (2.52)} & {\tiny 0.8$\pm$0.1}   & {\tiny 1.4 (2.3)}   & {\tiny 3.5 (4.0)}   & {\tiny 0.45 (0.25)} & {\tiny -2.54 (-3.05)} & {\tiny  1.79 (1.66)} \\    
  \hline   
  {\tiny $\langle$Corrected$\rangle$} & & & {\tiny {\bf 44$\pm$7}} & {\tiny {\bf 2.42$\pm$0.38}} & & {\tiny {\bf 1.88$\pm$0.26}} & {\tiny {\bf 4.70$\pm$0.42}} & {\tiny {\bf 0.42$\pm$0.04}} & {\tiny {\bf -2.55$\pm$0.49}} & {\tiny {\bf 1.90$\pm$0.22}} \\  
  {\tiny $\langle$Original$\rangle$} & & & {\tiny 82$\pm$14} & {\tiny 4.42$\pm$0.77} & & {\tiny 3.85$\pm$0.55} & {\tiny 6.83$\pm$0.95} & {\tiny 0.23$\pm$0.02} & {\tiny -3.22$\pm$0.74} & {\tiny 1.74$\pm$0.18} \\   
\end{tabular}    
\end{center}    
\vskip0.1cm    
\noindent (1) Distance D of the cloud. \\  
(2) Mass M $\propto N(H_2) D^2$. The H$_2$ column density determination assumes a   
mean atomic weight per molecule of 2.3. The total mass was determined above an \av\ level of $\sim$1 mag,   
which is a typical value for estimating molecular cloud extent (e.g., Lada et al. \cite{lada2010}).   
For the uncorrected maps, we determine the mass within the same area  
(above a threshold of \av\ = 1 + $\Delta$A$_{\rm v}$ [mag]). We note that for Auriga, the mass estimated by  
Harvey et al. (2013) is slightly higher (4.9 10$^4$ M$_\odot$) than our value. Both mass values were derived from  
an area of $\sim$13 deg$^2$, while the mass of Auriga given in Lada et al. (2010) of 1.1 10$^5$ M$_\odot$  
is determined from the whole cloud complex, covering  80 deg$^2$. \\    
(3) Surface density ($\Sigma$ = M/area). \\  
(4) Average column density (above a level of 10$^{21}$ cm$^{-2}$). \\  
(5) Background/foreground level of visual extinction. The error is the root mean square from the pixel statistics    
used to determine the contamination. \\  
(6) Peak of PDF in visual extinction. \\  
(7) Visual extinction value of the deviation point where the PDF starts to deviate from the lognormal    
shape at high column densities. \\     
(8) Dispersion of the fitted lognormal PDF. \\  
(9) Slope $s$ of the high-density tail of the PDF, determined by linear regression (the $\chi^2$ value is given in   
the panels of the PDF). We excluded the noisier and less well sampled points at the high column density end of the PDF. \\   
(10) Exponent of the spherical density distribution $\rho \propto r^{-\alpha}$, determined from $s$ with   
$\alpha = -2/s+1$ (Federrath \& Klessen \cite{fed2013}).   
\end{table*}   
 
\section{Introduction} \label{intro} Recent years have seen
significant progress in understanding the link between the column
density and spatial structure of molecular clouds and star formation.
Molecular line surveys enabled us to study the velocity structure of
clouds and yielded substantial results.  Just to name a few, the very
complex velocity structure of filamentary clouds was shown in a large
Taurus survey (Goldsmith et al.  \cite{goldsmith2008}, Hacar et al.
\cite{hacar2013}), the influence of large-scale convergent flows as a
possible formation mechanism for molecular clouds was demonstrated by
Motte et al.  (2014), and the importance of global collapse of
filaments for the formation of OB clusters was shown in Schneider et
al. (2010).  However, molecular tracers are restrained by their
critical densities and optical depths effects, thus limiting their
functionality to study only certain gas phases (for example, low-J CO
lines for low-density gas or N$_2$H$^+$ for cold, high-density gas).
On the other hand, widefield extinction maps obtained by near-IR
color-excess techniques cover a wider range of column densities,
typically from N(H+H$_2$) $\sim$0.1--40 10$^{21}$ cm$^{-2}$ (Lada et
al.  \cite{lada1994}; Lombardi \& Alves \cite{lombardi2001}; Lombardi
\cite{lombardi2009}; Cambr\'esy et al.  2011, 2013; Rowles \&
Froebrich \cite{rowles2009}; Schneider et al.  \cite{schneider2011}).
Some major results obtained from these studies are, for example, that
there is a relation between the rate of star formation and the amount
of dense gas in molecular clouds (Lada et al.  \cite{lada2010}), that
a universal threshold in visual extinction of \av\ $\approx$6 mag for
the formation of stars could exist (Froebrich \& Rowles
\cite{froebrich2010}), and that there are characteristic size scales
in molecular clouds, indicating the dissipation and injection scales
of turbulence (Schneider et al.  \cite{schneider2011}).
 
The large-scale far-infrared dust emission photometric observations of
{\sl Herschel}\footnote{Herschel is an ESA space observatory with
  science instruments provided by European-led Principal Investigator
  consortia and with important participation from NASA.} (Pilbratt et
al. \cite{pilbratt2010}) allow us now to make column density maps with
a very wide dynamic range N(H$_2$)$\sim$10$^{20}$ cm$^{-2}$ to a few
hundred 10$^{23}$ cm$^{-2}$ at an angular resolution of $\sim$25$''$
to $\sim$36$''$ that provide an exceptional database to better
understand the composition and structure of the interstellar medium.
One result of {\sl Herschel} is the importance of filaments for the
star-formation process. Though the filamentary structure of molecular
clouds has always been recognized, only the detailed investigation of
the column density structure of filaments (Molinari et al.
\cite{molinari2010}, Arzoumanian et al.  \cite{doris2011}, Palmeirim
et al. \cite{pedro2013}), their link to core formation (e.g., Andr\'e
et al.  \cite{andre2010}, \cite{andre2014}, Polychroni et al.
\cite{poly2013}, K\"onyves et al., in prep.), and their high mass
input to form OB clusters (Schneider et al.  \cite{schneider2012},
Hennemann et al.  \cite{hennemann2012}) emphasized their role for the
formation of stars.
 
Despite this progress, there are still a number of important questions
that are open.  1. What is the {\sl \emph{relative importance of
    turbulence, gravity, magnetic fields, and radiative feedback}} for
regulating the overall column density structure of molecular clouds?
2. Are there differences in the column density structure of clouds
forming \emph{\emph{{\sl \emph{low-mass stars or high-mass stars}}}}?
3. Is there a {\sl \emph{universal (column) density threshold}} for
the formation of self-gravitating prestellar cores? 4. Does the\emph{
  {\sl \emph{star formation efficiency}}} (SFE) and \emph{{\sl
    \emph{star-formation rate}}} (SFR) depend on the column density
structure of molecular clouds?
  
In a series of papers using column density maps obtained with {\sl
  Herschel} data and with near-IR extinction and the results of
numerical simulations, we address these questions.  This paper makes a
start with a detailed study of the validity of column density maps and
their probability distribution functions (PDFs) obtained from {\sl
  Herschel}.  We show that line-of-sight (LOS) confusion, i.e.,
emission from diffuse dust mixed with low-density gas as well as
denser clouds in front or behind the bulk emission of the molecular
cloud, leads to a significant overestimation in the column density
maps. Apart from this observational approach, we quantify how the PDF
properties change by using numerical simulations in which we add noise
and foreground and/or background emission to an uncontaminated PDF.
Using the corrected maps, we then study their PDFs and column density
profiles to address the question of whether all molecular clouds have
a similar column density structure.
 
\section{Line-of-sight confusion in continuum maps}  
 
A notorious problem of continuum maps (extinction maps or column
density maps derived from SED fitting using {\sl Herschel}) is
line-of-sight confusion. This is particularly true in the Galactic
plane and along spiral arms where dust emission not related to the
cloud can significantly add, hence artifically increase, the derived
column density for the cloud studied at a single distance.  However,
it is not easy to distinguish between the bulk emission of the cloud
and background or foreground contribution from unrelated clouds if the
distance information is missing.  A simple approach we present in this
paper is to determine a mean value for the contaminating column
density by averaging pixels outside the molecular cloud close to the
map borders. In the figures of column densities shown in the appendix,
this area is outlined as a white, dashed polygon. We find values
between \dav\ $\sim$0.8 and 3 (see Table~\ref{table:summary}), with a
standard deviation of 0.1 to 0.5, which are then subtracted as an
offset from the original column density map.  Such a constant offset
subtraction is also intrinsically a part of the NICER/NICEST
extinction mapping techniques (see Lombardi \& Alves (2001) and
Lombardi et al. (\cite{lombardi2008}); however, major uncertainty in
the background determination in our method comes from the placement of
the field and not so much from the variation within.  It can be that
the `reference field' is still too close to the cloud itself and may
contain contributions from the cloud. We exclude this effect for
Carina and Maddalena by comparing to the near-IR extinction maps
(Bontemps, priv. comm, see Schneider et al.  2011 for the method) of
the large-scale environment of the clouds.  For Auriga, we are also
well outside of the cloud, based on the extinction map shown in Lada
et al. (2009). It is only for NGC3603 that such a selection bias could
potentially be a problem.
 
We emphasize that our approach is simple and ignores possible
gradients and more generally the spatial variation of the emission
distribution of clouds that are located along the line of sight. (More
distant background clouds are expected to be smaller than foreground
clouds that are assumed to be bigger in angular size.)  This can lead
to unrealistic features in the PDF (see Sect.~\ref{theory}).  For
simplicity, we only consider correcting for foreground/background
contamination here by subtracting a constant value, which already
provides significantly more reliable estimates of the true column
density and masses.
  
An alternative method of estimating the contamination is to use the
velocity information from atomic hydrogen or molecular lines and to
translate the observed intensity into H$_2$+HI column density. This
has been done for {\sl Herschel} studies of NGC6334 (Russeil et al.
2013) and W3 (Rivera-Ingraham et al.  \cite{alana2013}).  However,
using $^{12}$CO or $^{13}$CO is of very limited use for more distant
and massive clouds because these tracers are only sensitive up to an
\av\ of typically 10 mag.  Probability distribution functions of CO
(Goldsmith et al. \cite{goldsmith2008}, Goodman et al.
\cite{good2009}) are clipped at higher column densities owing to a
high optical depth.  Only recently has a special method been developed
to obtain an H$_2$ map from $^{13}$CO and C$^{18}$O observations in
W43 that suffers less from a cut-off at high column densities
(Carlhoff et al. \cite{carlhoff2013}).  However, all methods are still
subject to other uncertainties, such as the variable CO/H$_2$
conversion factor or the variation in excitation temperature T$_{ex}$
(see Fig.~11 in Carlhoff et al. \cite{carlhoff2013} that shows how the
PDF shifts by changing only a few Kelvin T$_{ex}$).
 
\section{Probability distribution functions (PDFs) of column density}   
 
\subsection{PDFs in theory} \label{pdf-simu} 
 
The PDF of column density $N$ is defined as the probability of finding
gas within a range [$N$, $N$+d$N$].  The column density $N$ can be
replaced by the visual extinction \av\footnote{N(H$_2$) = \av\
  0.94$\times$10$^{21}$ cm$^{-2}$ mag$^{-1}$ (Bohlin et al.
  \cite{bohlin1978})}. We define $\eta\equiv\rm \ln(N/ \langle N
\rangle)$ as the natural logarithm of the column density N, divided by
the mean column density $\langle N \rangle$, and the quantity
$p_\eta(\eta)$ then corresponds to the PDF of $\eta$ with the
normalization $\int_{-\infty}^{+\infty} p_\eta \rm d \eta =
\int_0^{+\infty} p_N\, dN=1$.
 
Volume density PDFs are an important tool for characterizing the
properties of the interstellar medium (e.g., Padoan et al.
\cite{padoan1997}; Vazquez-Semadeni \& Garcia \cite{vaz2001}; Burkhart
et al., \cite{burk2013}); models of star formation, in particular the
stellar initial mass function (e.g.,\, Hennebelle \& Chabrier 2008,
2009; Padoan \& Nordlund \cite{padoan2002}); and the star formation
rate (Krumholz \& McKee \cite{krumholz2005}, Federrath \& Klessen
\cite{fed2012}).  However, the 3D-density distribution in a molecular
cloud cannot be directly observed; instead, only the 2D-projection,
i.e. the column density field, is accessible.  Numerical simulations
show that the column density PDFs have a smaller width than the
density PDFs and can show slightly different shapes in the high- and
low-density regimes (Federrath et al.  \cite{fed2010}), but they are
statistically equivalent (Federrath \& Klessen \cite{fed2013}).  Any
particular, single column density projection can deviate from the
volume density PDF, but the column and volume density PDFs show the
same features if isotropy is assumed and/or if averaged over many
realizations and/or different projections (Vazquez-Semadeni \& Garcia
\cite{vaz2001}, Fischera \& Dopita \cite{fischera2004}, Federrath et
al.  \cite{fed2010}).  A method using the column density power
spectrum has been developed by Brunt et al.  (\cite{brunt2010}) to
approximate the 3D-PDF from a 2D-PDF.
 
\begin{figure*}[!htpb]     
\centering    
\includegraphics [width=8cm, angle={0}]{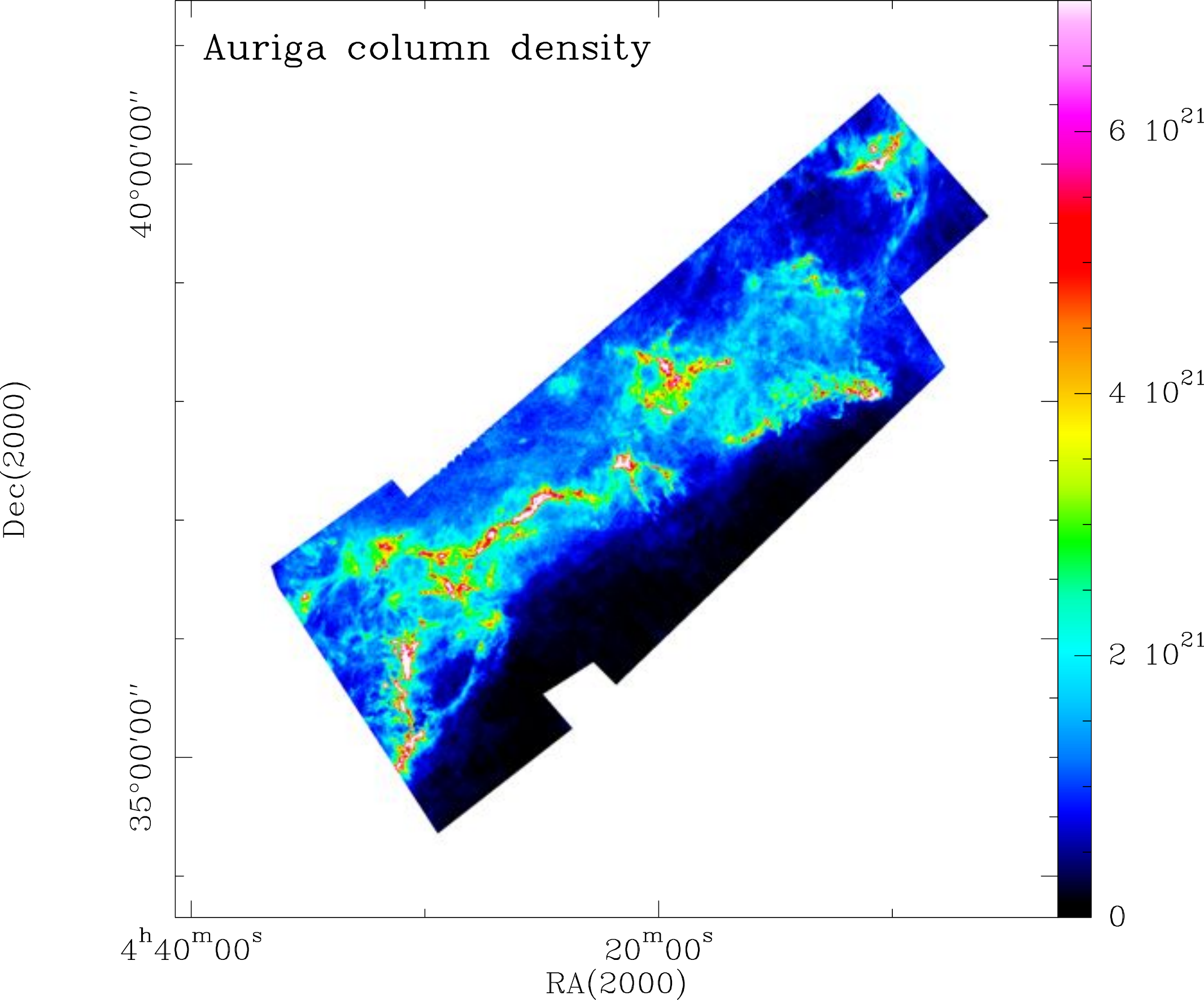}    
\includegraphics [width=8cm, angle={0}]{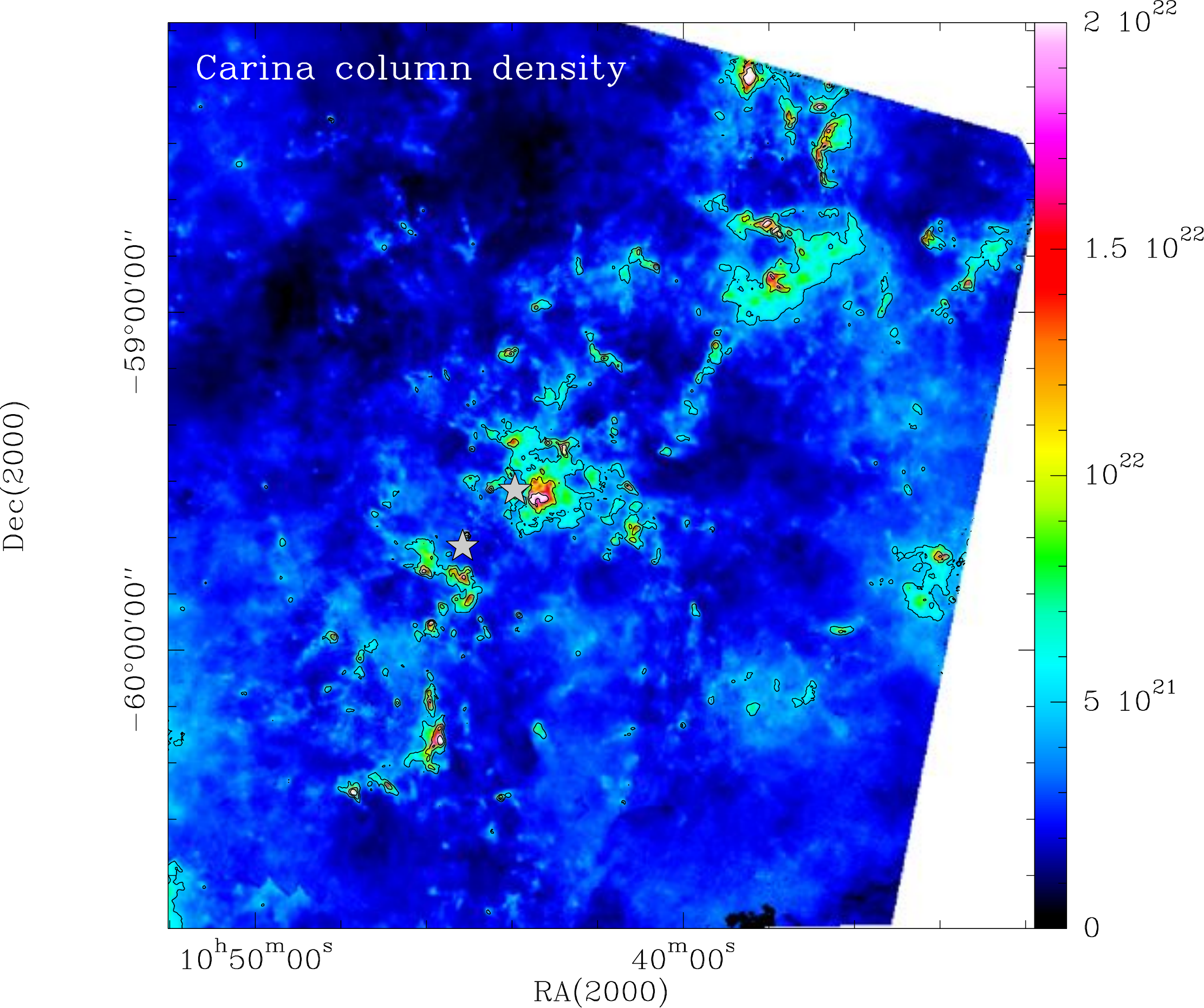}    
\includegraphics [width=8cm, angle={0}]{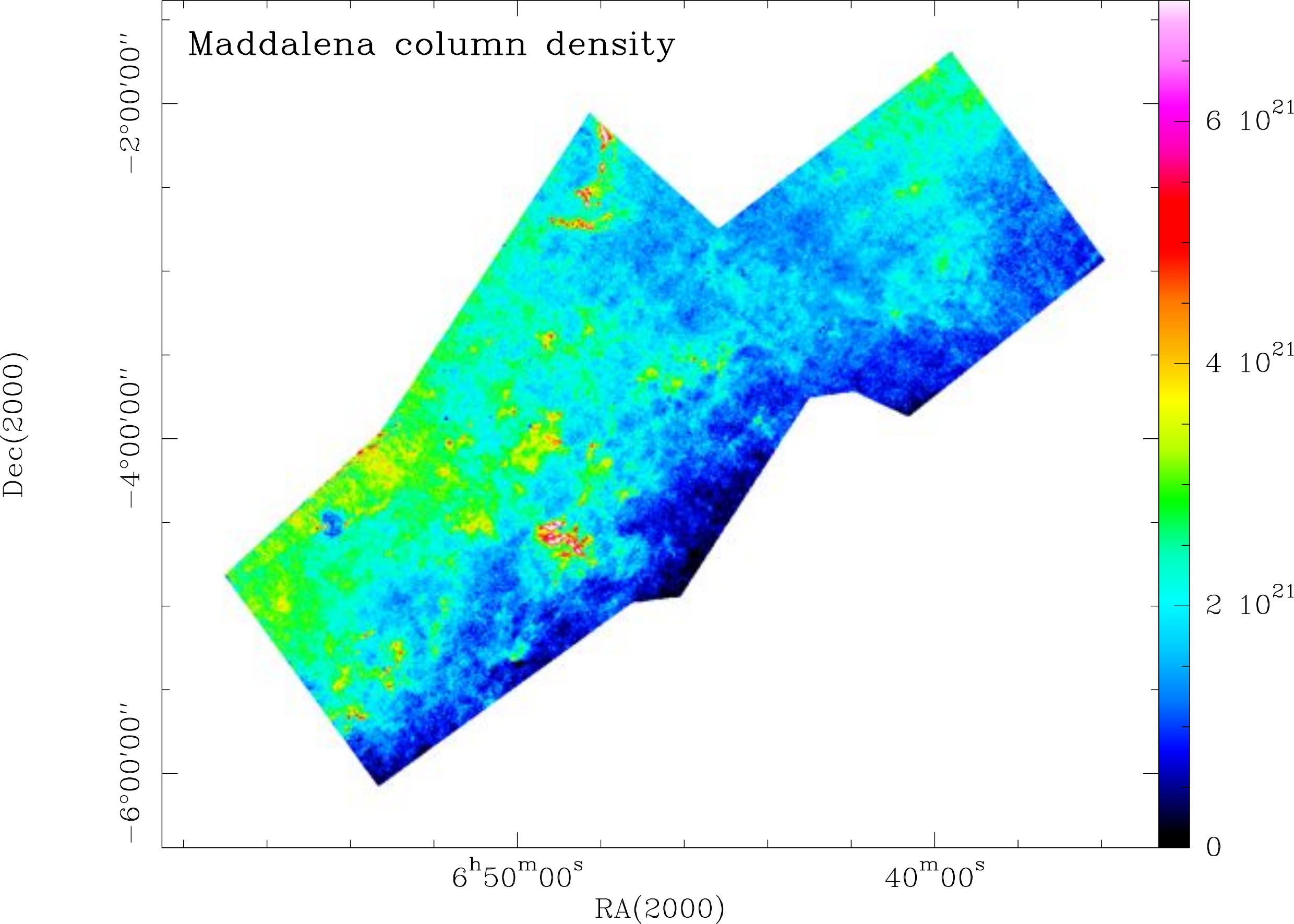}    
\includegraphics [width=8cm, angle={0}]{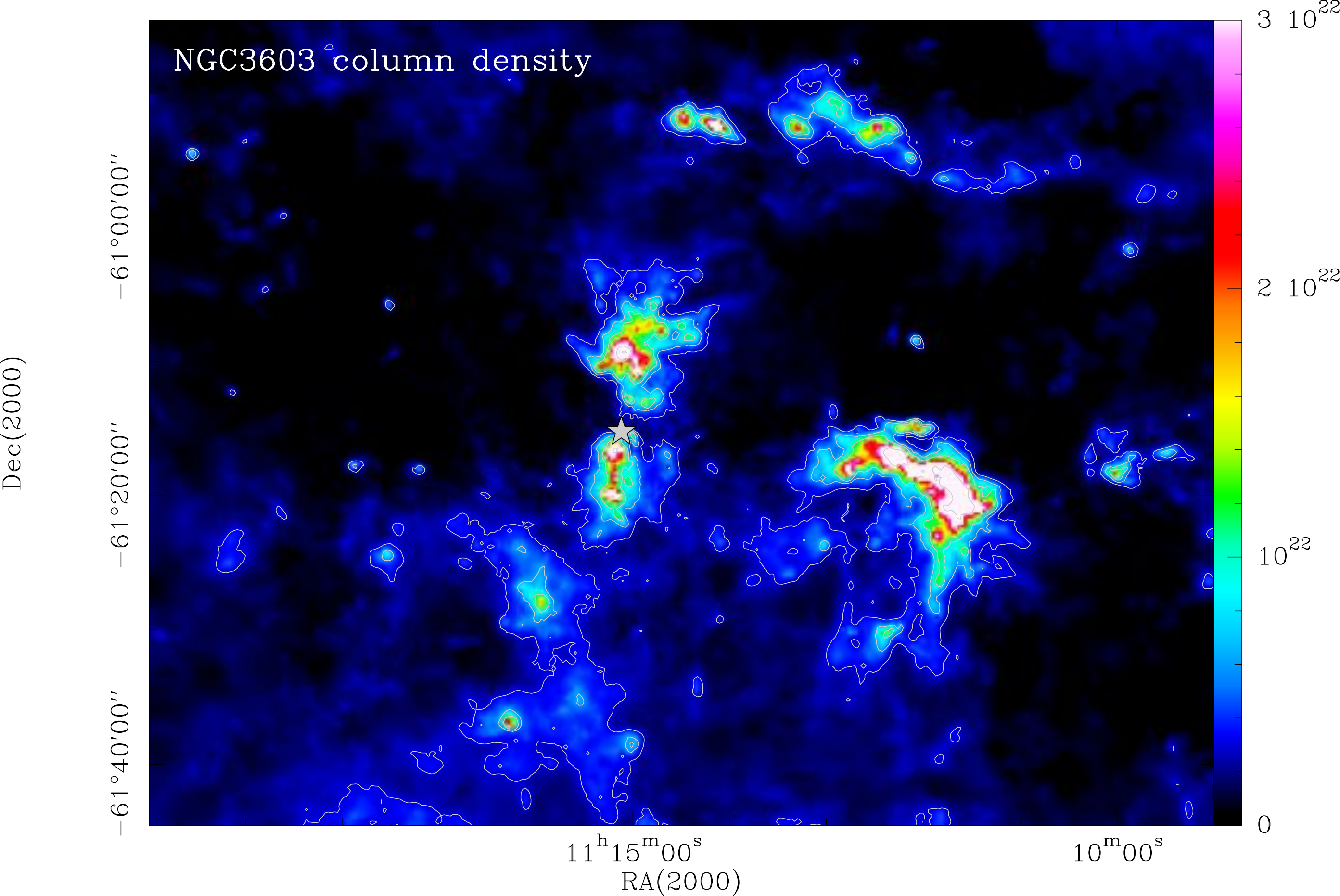}    
\caption[] {{\sl Herschel} column density maps (in [cm$^{-2}$], all starting at zero)  
of Auriga, Maddalena, NGC3603, and Carina after correcting for line-of-sight contamination  
and removing noisy edges and areas where there was no overlap between PACS and SPIRE. 
The contour levels are 3, 6, 10, and 50 10$^{21}$ cm$^{-2}$ for NGC3606 and  
5, 10, and 20 10$^{21}$ cm$^{-2}$ for Carina. }   
\label{cds}    
\end{figure*}    
 
Generally, the shape and dispersion of the PDF are determined by the
density variations in a turbulent medium due to the contribution of
compressive forcing, the velocity structure (expressed by the sonic
Mach number), and magnetic pressure (e.g., Molina et al.
\cite{molina2012}, Federrath \& Klessen \cite{fed2012}).  From a
theoretical point of view, the PDF of a molecular cloud shows a
lognormal distribution if the cloud structure is governed by
isothermal supersonic turbulence alone and all turbulence modes are
sufficiently statistically sampled in the observed density structure.
If there are statistical density fluctuations and/or intermittency due
to locally compressive turbulence, some deviations from the lognormal
shape at higher column densities can be observed as seen in the PDF of
Polaris (Schneider et al. 2013).  Significant deviations are predicted
when parts of the gas undergo gravitational collapse and are thus
governed by their self-gravity (Klessen 2000, Federrath et al.  2008,
Kritsuk et al.  \cite{kritsuk2011}, Ward et al. 2014).  Passot \&
Vazquez-Semadeni (\cite{passot1998}) showed that for nonisothermal
gas, a power-law regime forms as well.  Depending on the polytropic
exponent $\gamma_p$ in the pressure term $P \propto \rho^{\gamma_p}$
with density $\rho$, a power-law tail forms in the case of $\gamma_p <
1$ for high densities and $\gamma_p > 1$ for low densities. Several
authors (Ballesteros-Paredes et al.  \cite{ball2011}, Cho \& Kim
\cite{cho2011}, Ward et al. 2014) indicated that PDFs from
hydrodynamic simulations including gravity change during the cloud
evolution, and Federrath \& Klessen (\cite{fed2013}) showed in their
models how the slope of the high-density tail of the PDF flattens with
increasing star-formation efficiency. Recently, Girichidis et al.
(\cite{giri2014}) have demonstrated analytically that free-fall
contraction of a single core or an ensemble of collapsing spheres
forms a power-law tail.
     
\subsection{PDFs from observations}  \label{pdf-obs} 
 
To observationally determine PDFs, the most common method so far has
been to employ column density maps obtained from near-IR extinction
(see Sec.~\ref{intro}) and to derive a normalized distribution of
pixels vs.  column density in logarithmic bins.  The observed PDFs
show a lognormal distribution over a range of low column densities,
typically between \av $\sim$0.5 to $\sim$4.  A more or less lognormal
distribution was reported for clouds (Lupus, Coalsack: Kainulainen et
al. 2009; Polaris: Schneider et al. 2013; Chamaeleon III: Alves de
Oliveira et al.  \cite{catarina2014}) where none or only a few pre- or
protostellar cores were detected. {\sl Herschel} studies of several
star-forming molecular clouds (Tremblin et al.  \cite{tremblin2014};
Schneider et al. 2012, 2013) have shown that the lognormal part of the
PDF broadens in the presence of external pressure (e.g., expanding
\hii\ regions) and that even a second peak -- indicating the
compressed shell -- can appear. Deviations from the lognormal shape in
the form of a power-law tail are found for low-to-intermediate-mass
star-forming regions (Lombardi et al.  \cite{lombardi2008}), where it
was proposed that the excess at high column densities is related
either to star-formation activity (Kainulainen et al.  2009) or to the
pressure due to different phases in the interstellar medium
(Kainulainen et al.  2011).  Gravity as the clearly dominating (over
pressure and magnetic fields) process to form a power-law tail was
suggested by Froebrich \& Rowles (\cite{froebrich2010}) and Schneider
et al.  (\cite{schneider2013}).
 
We here determine PDFs from {\sl Herschel} column density maps on a 
14$''$ grid (see Appendix B).  The maps are on average large 
($>$1$^\circ$) and have a size of a few hundred$^2$ pixels up to a few 
thousand$^2$ pixels. We choose a binsize of 0.1 in $\eta$ that 
provides the best compromise between resolving small features in the 
PDF and avoiding low pixel statistics. In Appendix A, we show that 
binning in smaller or larger bins or varing the pixelsize and resolution 
does not change the PDF properties (width, peak, slope of power-law 
tail). To derive the characteristic properties of the PDF 
(width $\sigma_\eta$, peak A$_{\rm v,pk}$, deviation from the 
lognormal shape \av(DP)), we fit the lognormal function 
\begin{equation}    
p_\eta\,{\rm d}\eta=\frac{1}{\sqrt{2\pi\sigma_\eta^2}}{\rm exp}\Big[ -\frac{(\eta-\mu)^2}{2\sigma_\eta^2} \Big]{\rm d}\eta    
\end{equation}    
where $\sigma_\eta$ is the dispersion and $\mu$  the mean 
logarithmic column density. We do this systematically by performing 
several fits on a grid of parameters for $\eta$ and $\mu$ and then 
calculating the positive and negative residuals. Since the power-law tail 
is expected to cover higher $\eta$ values than the lognormal part, we 
select fits with the least negative residuals in the fitting 
procedure. We then determine the range of lognormality, when the 
difference between the model PDF (eq. (1)), and the measured PDF 
$p_\eta$ is less than three times the statistical noise in $p_\eta$. 
    
As shown by various authors (e.g., Kainulainen et al.  \cite{kai2009},
Hill et al. \cite{hill2011}, Russeil et al. 2013, Schneider et al.
2012, 2013), a power-law tail for high column densities emerges for
star-forming regions. We perform a linear regression fit in order to
determine the slope $s$ of the power-law tail. The values taken into
account start at the deviation point (DP) where the lognormal PDF
turns into a power-law distribution and stop where the power law is no
longer well defined (at high column densities)because of the lower
pixel statistics caused by resolution effects. In case the tail is
only due to gravity, and if we assume spherical symmetry, the PDF
slope $s$ of the power-law tail is related to the exponent $\alpha$ of
the radial density profile (Federrath \& Klessen \cite{fed2013}) with
\begin{equation}   
\rho  \propto r^{-\alpha}   
\end{equation}   
and 
\begin{equation}   
\alpha = -2/s +1 
.\end{equation} 
  
\begin{figure*}[!htpb]    
\centering    
\vspace{0cm}\includegraphics [width=7.5cm, angle={90}]{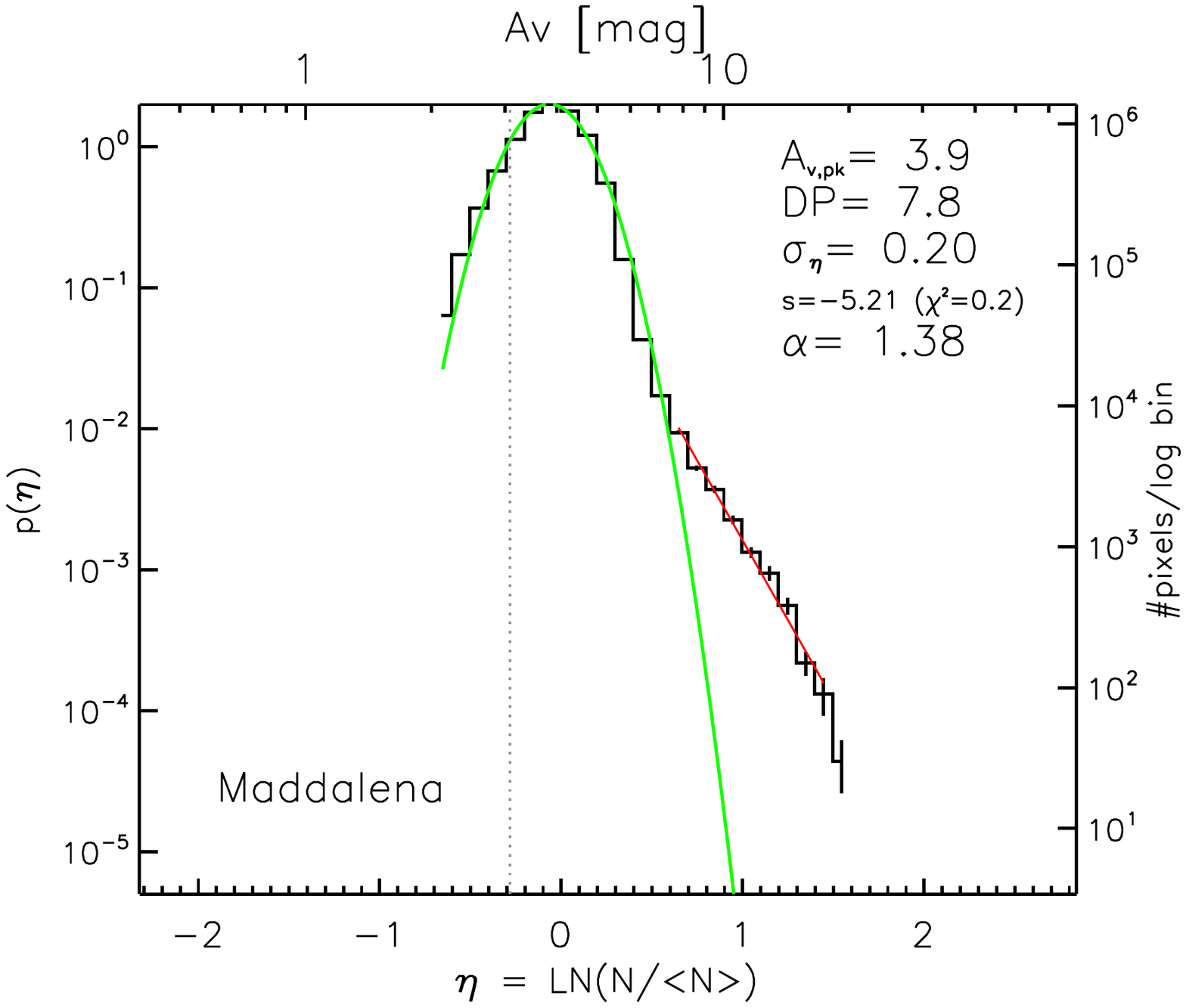}    
\hspace{-1.5cm}\includegraphics [width=7.5cm, angle={90}]{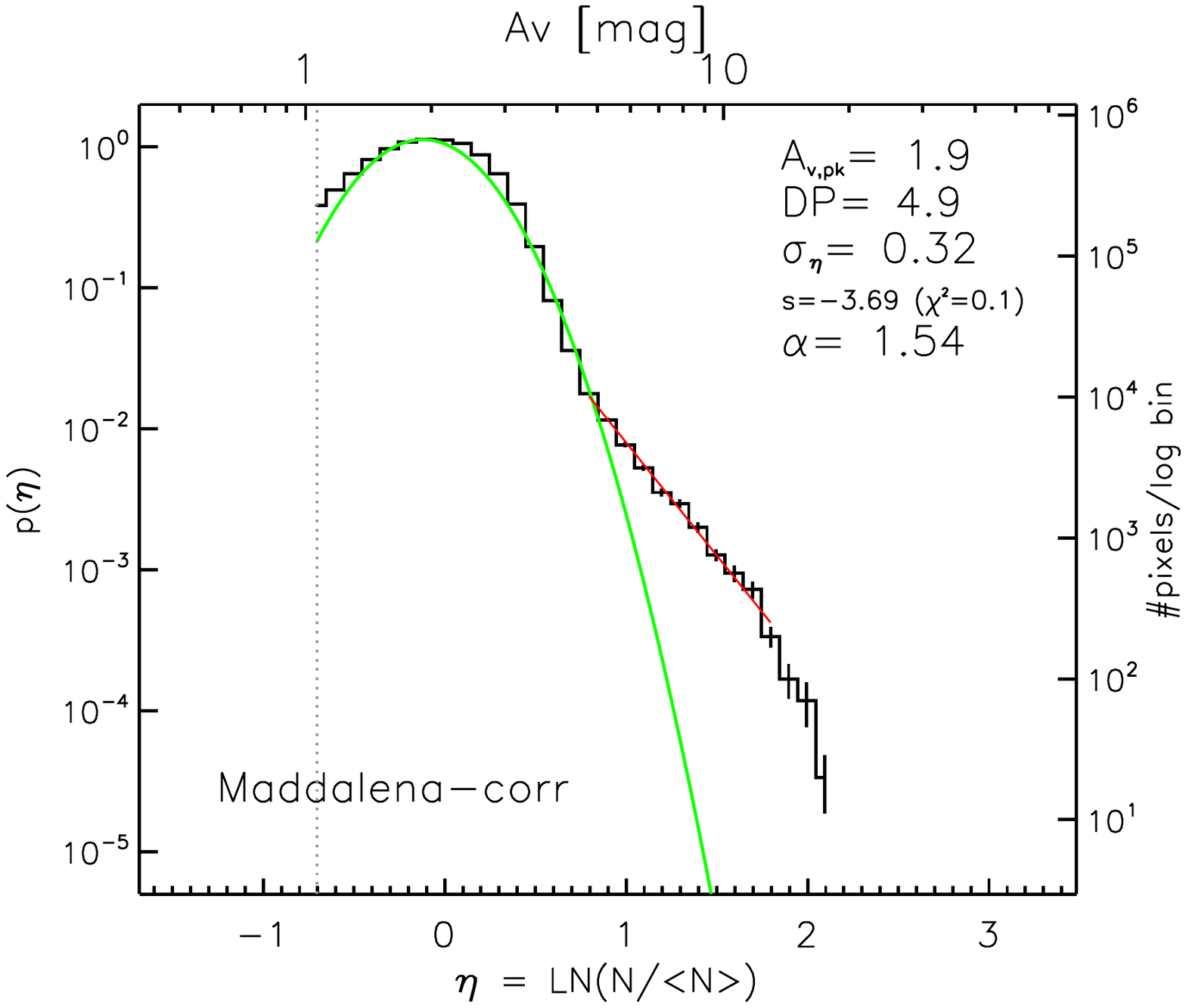}    
  
\vspace{-2.0cm}  
\hspace{0cm}\includegraphics [width=7.5cm, angle={90}]{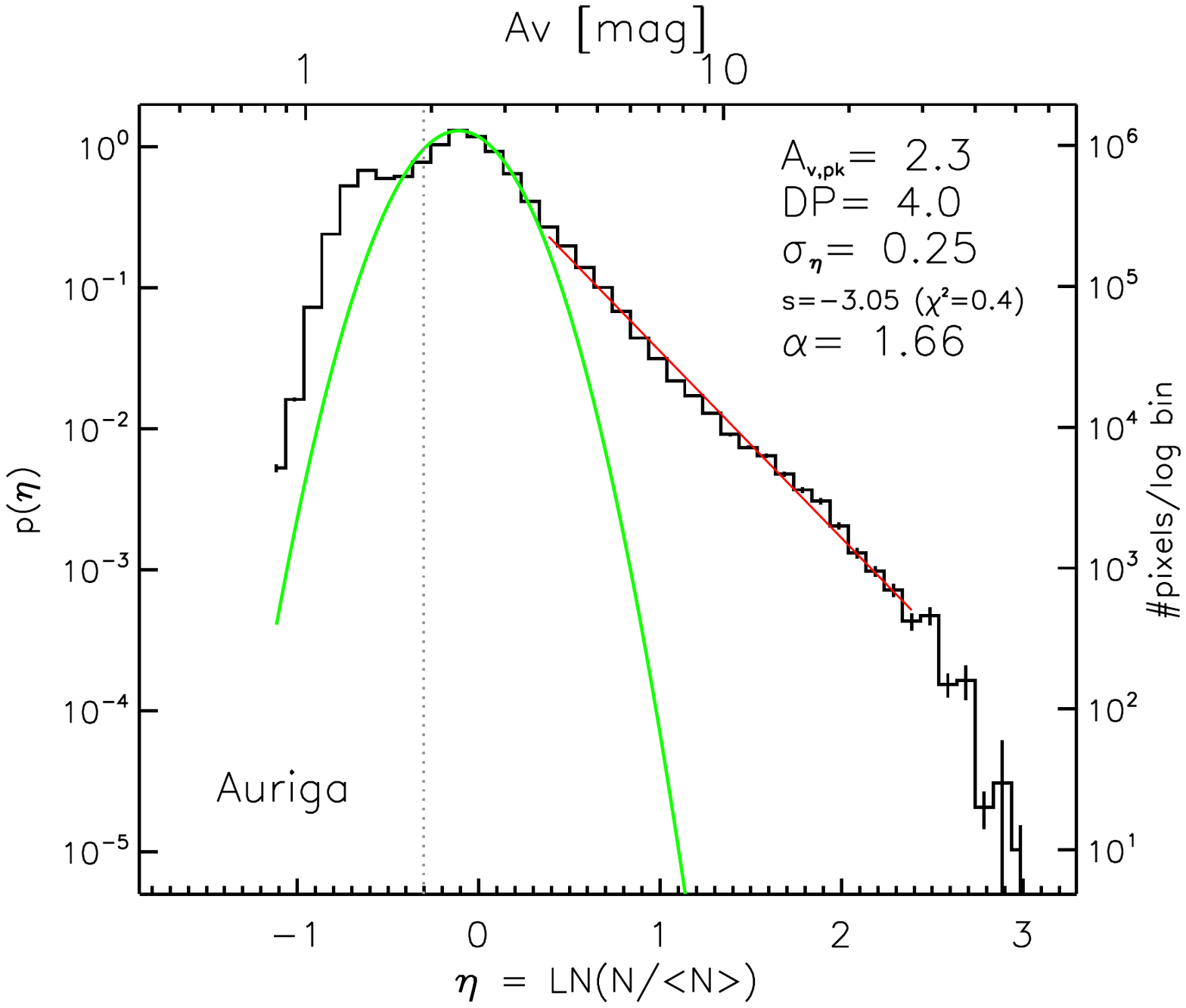}    
\hspace{-1.5cm}\includegraphics [width=7.5cm, angle={90}]{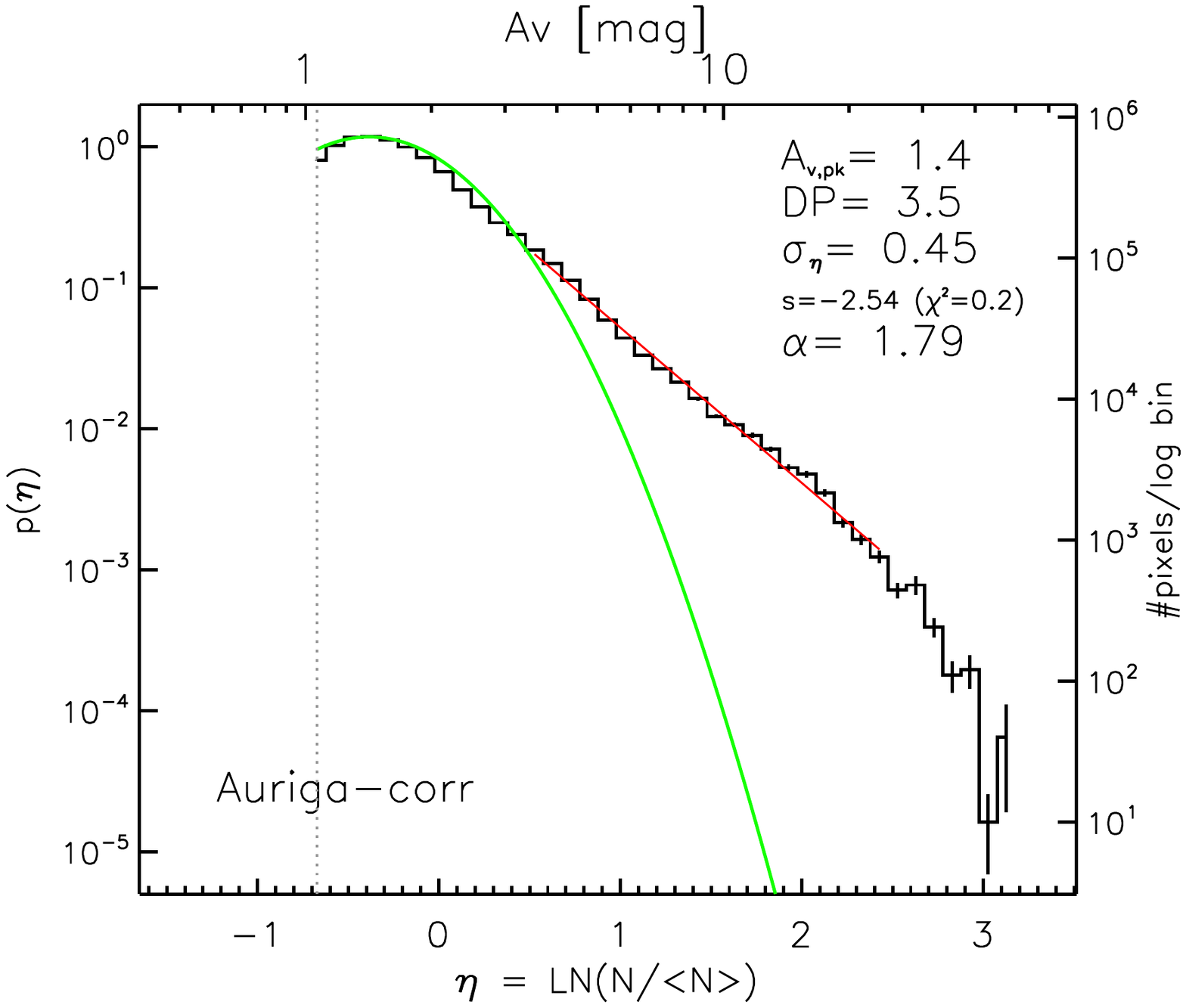}    
\caption[] {PDFs derived from {\sl Herschel} column density maps 
  toward the {\bf low-mass star-forming regions} Maddalena and 
  Auriga.  The left panel shows the PDF from the original map, the 
  right one from the corrected map. The vertical dashed line indicates 
  the approximate completeness limit.  The left y-axis gives the 
  normalized probability $p(\eta)$, the right y-axis the number of 
  pixels per log bin. The upper x-axis is the visual extinction and 
  the lower x-axis the logarithm of the normalized column density. 
  The green curve indicates the fitted PDF.  For Auriga, diffuse 
  LOS-contamination (or a well defined seperate cloud) shows up as an 
  individual PDF at low extinctions.  The red line indicates a 
  power-law fit to the high \av\, tail.  Inside each panel, we give 
  the value where the PDF peaks (A$_{{\rm v},pk}$), the deviation 
  point from lognormal to power-law tail (DP), the dispersion of the 
  fitted PDF ($\sigma_\eta$), the slope $s$ and the $X^2$ of the fit 
  (linear regression), and the exponent $\alpha$ of an equivalent 
  spherical density distribution. These values are also summarized in 
  Table~1.} 
\label{fig:pdf-low}    
\end{figure*}

\begin{figure*}[!htpb]    
\centering    
\includegraphics [width=7.5cm, angle={90}]{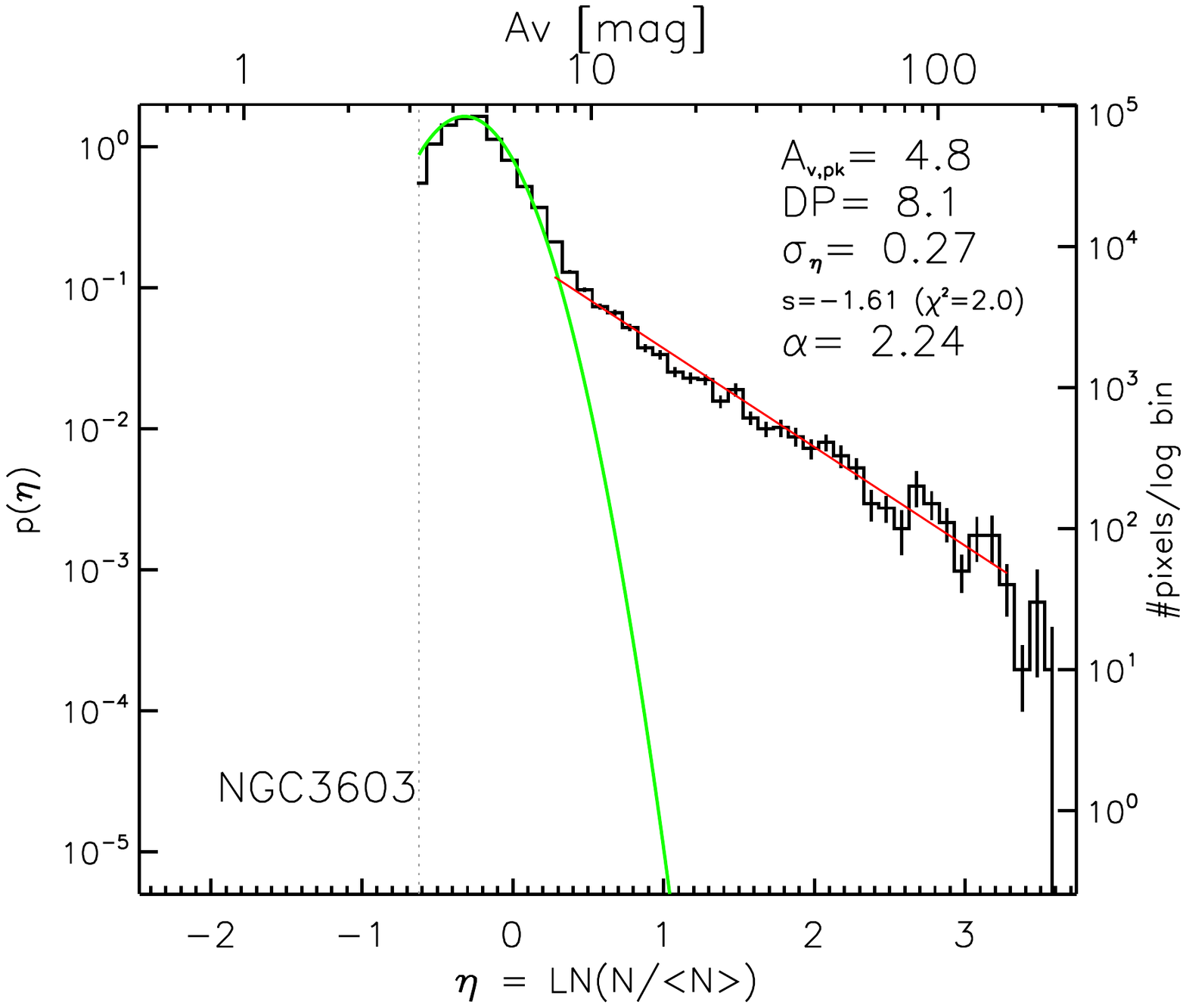}    
\hspace{-1.5cm}\includegraphics [width=7.5cm, angle={90}]{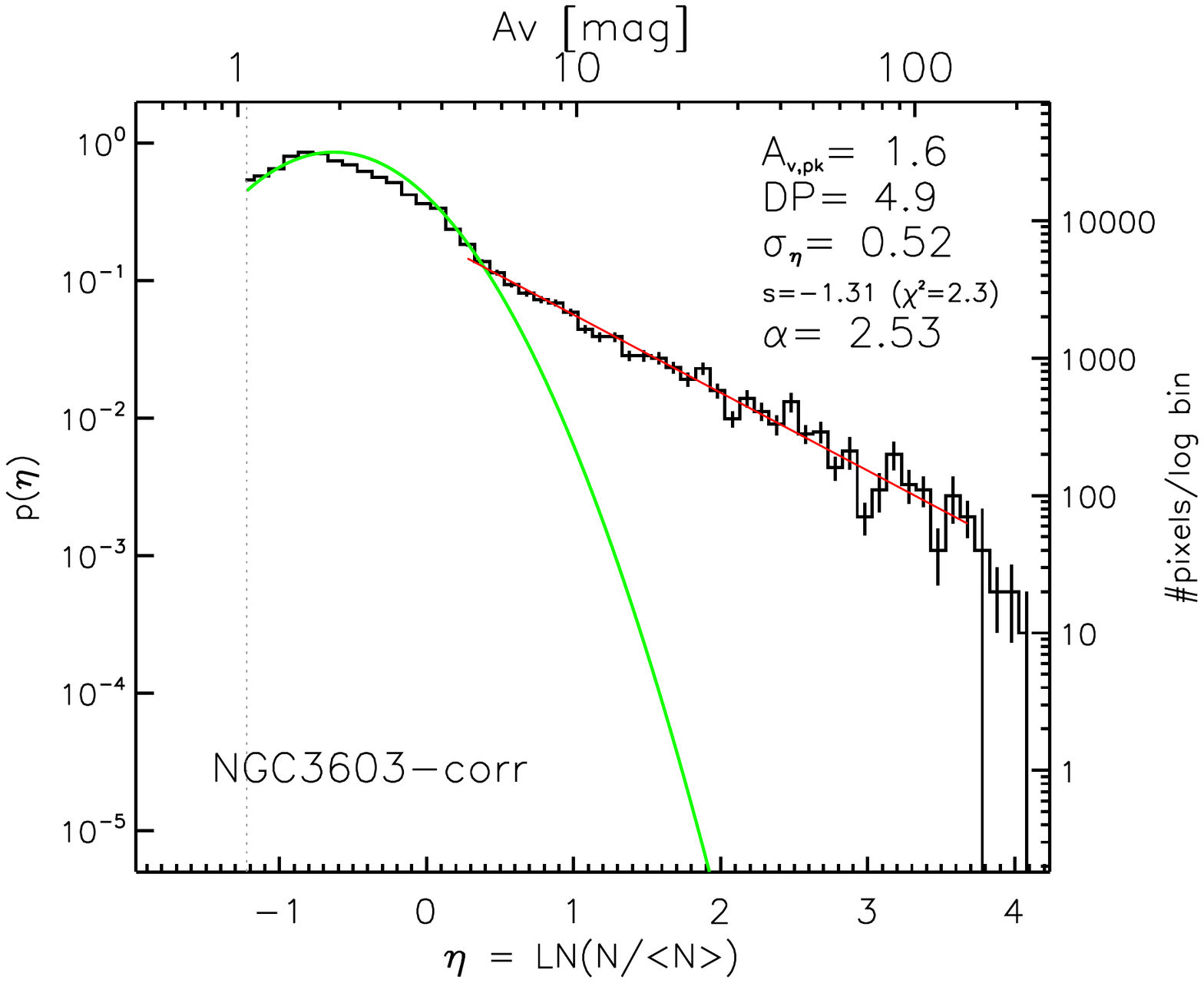}    
   
\vspace{-2cm}\includegraphics [width=7.5cm, angle={90}]{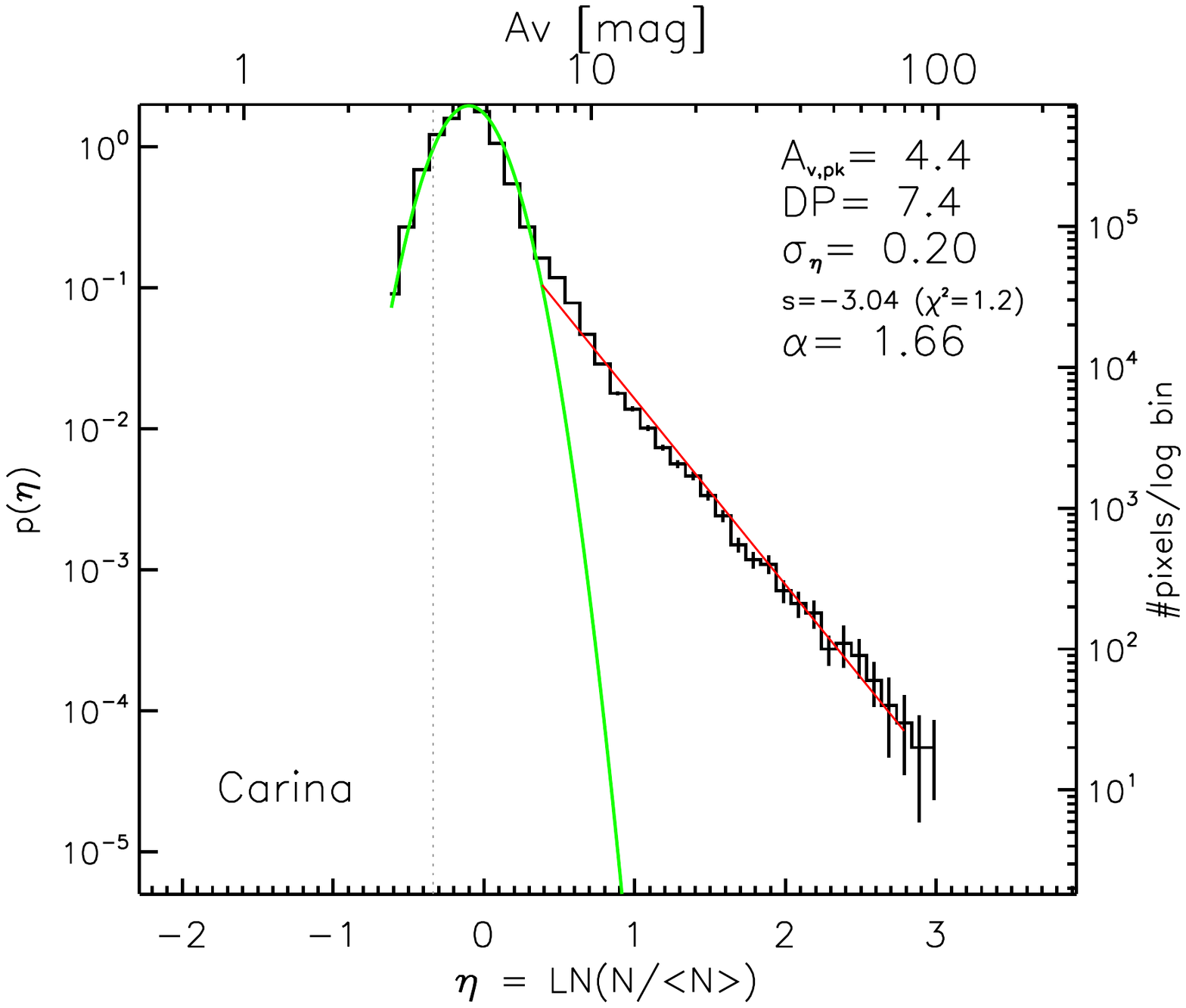}    
\hspace{-1.5cm}\includegraphics [width=7.5cm, angle={90}]{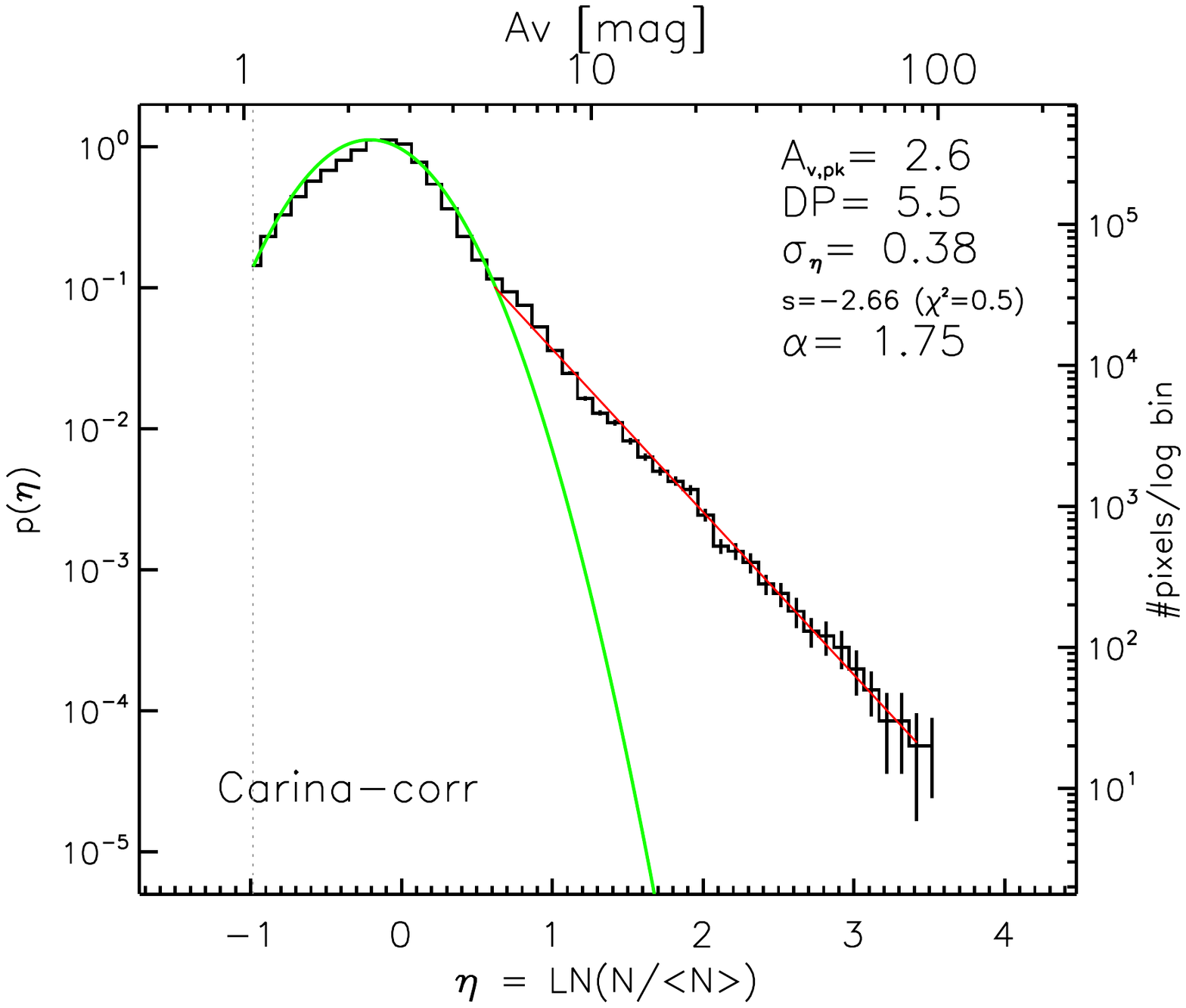}    
   
\caption[] {PDFs derived from {\sl Herschel} column density maps  
  toward the {\bf high-mass star-forming regions} NGC3603 and Carina.  
  The left panel shows the PDF from the original map, the right one  
  from the corrected map. All other parameters as in Fig.~1.}  
\label{fig:pdf-gmc}    
\end{figure*}    
  
\section{Herschel column density maps} \label{obs} For our study, we 
obtained the raw data from the {\sl Herschel} archive for the 
following sources (see Table~\ref{table:summary}): {\bf 
  Auriga-California:} RA(2000)=4$^h$21$^m$, Dec(2000)=37$^\circ$35$'$; 
l=162.3663$^\circ$, b=--8.7607$^\circ$, OT1 PI: P. Harvey, see Harvey 
et al.  (\cite{harvey2013}); {\bf Maddalena:} RA(2000)=6$^h$48$^m$, 
Dec(2000)=--3$^\circ$39$'$; l=215.7908$^\circ$, b=--2.4273$^\circ$, 
OT2 PI: J.  Kauffmann; {\bf NGC3603:} RA(2000)=11$^h$15$^m$, 
Dec(2000)=--61$^\circ$15$'$; l=291.6000$^\circ$, b=--0.5180$^\circ$, 
Hi-GAL (``{\it Herschel} Infrared GALactic plane survey'', Molinari et 
al.  (\cite{molinari2010}); {\bf Carina:} RA(2000)=10$^h$43$^m$, 
Dec(2000)=--59$^\circ$24$'$; l=287.2328$^\circ$, b=--0.5009$^\circ$, 
OT1 PI: T.  Preibisch, see Preibisch et al.  (\cite{preibisch2012}), 
Gaczkowski et al. (\cite{gac2012}), Roccatagliata et al. 
(\cite{roca2013}). These clouds were selected to cover 
different masses, sizes, and levels of SF.  The Maddalena cloud is 
very massive but shows only a low level of SF. Auriga is less massive 
but forms low-mass stars, and Carina and NGC3603 are high-mass SF 
regions with associated OB clusters. The process of data reduction and 
determination of column density maps at an angular resolution of 
$\sim$36$''$ is explained in Appendix B where we also show the 
resulting column density maps and individual 250 $\mu$m maps. 
 
\section{Column density maps and PDFs} \label{coldens}     
   
Figure~\ref{cds} shows the column density maps of the four clouds,
corrected for contamination that can directly be compared with the
original ones in the Appendix. (Both are plotted in the same column
density range starting at zero.)
 
For a correct interpretation of the PDFs, the (non)-completeness of
the map needs to be considered. Our {\sl Herschel} maps cover most of
the molecular cloud material but do not extend to the very low column
density outskirts of the cloud.  For example, the extinction map of
Auriga from Lada et al.  (\cite{lada2009}), covers $\sim$80 deg$^2$
including gas with \av\ $\lesssim$1, while our {\sl Herschel} map
comprises an area of $\sim$13 deg$^2$. The lowest continuous contour
that is about complete in this map corresponds roughly to the
completeness limit, i.e., \av\,$\sim$1 for the corrected Auriga map.
The limit determined this way is \av\,$\sim$0.9, 1.0, and 1.3 for
Maddalena, NGC3603, and Carina (all for the corrected maps), which
also corresponds to the typical value taken for defining molecular
cloud extent (e.g., Lada et al. \cite{lada2010}). For simplicity, we
take \av\,=1 for all clouds as the completeness limit, though a strict
upper limit is probably 30\%--50\% higher. For the uncorrected column
densities, the completeness level is then the value determined from
the corrected maps plus the contamination level. However, the
additional uncertainty of determining the back- and foreground
contribution means that this level is more uncertain.
 
We emphasize that even if we are not fully complete for the lowest
column densities, the higher statistics of these low column density
pixels does not change the PDF shape significantly as we show in
Appendix A. There, we plot the PDF of Auriga above different
thresholds of column density, starting with the noise level of
\av\,=0.2. In summary, incompleteness/cropping can remove parts of the
lognormal low-column density part of the PDF but it does not change
the slope of the power-law tail. Cropping can indeed become an
important concern when studying infrared dark clouds, because these
can be embedded in a dense molecular envelope (see Schneider et al.
2014 for examples).  If the PDF is only constructed from higher column
density pixels, such a PDF underestimates the true, underlying PDF and
only the power-law tail remains.  On the other hand, intentionally
selecting subfields in a map to construct PDFs can be useful for the
physical interpretation.  This was done in Hill et al.  (2011),
Schneider et al. (2012), Russeil et al. (2013), and Tremblin et al.
(2014) to seperate the effects of turbulence, gravity, and compression
by ionization on the column density structure of a cloud.
 
The LOS contamination correction significantly reduces (up to a factor
$\sim$2) the absolute values for average column density $\langle
N(H_2) \rangle$, mass, and surface density
(Table~\ref{table:summary}).  The total mass of the cloud from the
uncorrected maps was determined above an \av\,-level of $\sim$1 mag,
and for the uncorrected maps, we determined the mass within the same
area (above a threshold of \av\ = 1 + $\Delta$A$_{\rm v}$ [mag]).  The
values for $\langle N(H_2) \rangle$ range between 1.5 and 3.2
$\times$10$^{21}$ cm$^{-2}$ with an average value of
(2.4$\pm$0.4$)\times$10$^{21}$ cm$^{-2}$. The surface density $\Sigma$
varies between 28 for Auriga and 60 for NGC3603 with an average value
of $\Sigma$ = 44 M$_\odot$ pc$^{-2}$ (instead of 82 M$_\odot$
pc$^{-2}$ that was obtained from the original maps).\footnote{We use
  the threshold \av = 1 mag to `define' a molecular cloud. Taking a
  higher value such as \av = 2 mag, we obtain a variation in $\Sigma$
  between 47 and 90 M$_\odot$ pc$^{-2}$ with an average value of 63
  M$_\odot$ pc$^{-2}$.}  This value corresponds very well to the one
found by Heyer et al.  ({\cite{heyer2009}) with $\Sigma$ = 42
  M$_\odot$ pc$^{-2}$ from a sample of clouds ($>$250) from the FCRAO
  Galactic Ring Survey investigated with $^{13}$CO 1$\to$0 emission,
  and to the value $\Sigma$ = 41 M$_\odot$ pc$^{-2}$ obtained from
  extinction maps of five nearby clouds (Lombardi et al.
  \cite{lombardi2010}).  However, we note that our simple approach may
  still under- or overestimate the LOS contamination.  In a
  forthcoming paper (Paper III), we will present a study of more than
  20 clouds to reach greater statistical significance.
  
  The PDFs determined from the original and contamination corrected 
  maps are shown in Figs.~\ref{fig:pdf-low} and \ref{fig:pdf-gmc}. 
  The low column density regime of the uncorrected maps is limited by 
  noise and LOS contamination of the map (\dav\,-value listed in 
  Table~\ref{table:summary}), the vertical dashed line in 
  Figs.~\ref{fig:pdf-low} and \ref{fig:pdf-gmc} indicates the 
  approximate completeness level (see above).  For Auriga, 
  the pixels left of the completeness limit still sample low-column 
  density molecular cloud material, though we most likely miss the 
  larger extent of this cloud component.  We then fitted with a 
  lognormal function the lower extinction part of the PDF as explained 
  in Sec.~\ref{pdf-simu}.\footnote{We keep the classical 
    approach to fit one lognormal PDF to the low column density range 
    though other functional fits, such as a Gaussian, can be possible as 
    well (Alves et al.  \cite{alves2014}).}  For the LOS-corrected 
  maps, the low extinction values (left of the PDF peak) do not 
  neccessarily have a perfect lognormal distribution (e.g., Maddalena) 
  because removing a constant offset can lead to negative pixels in 
  the maps that are ignored during the process to make the PDF.  
 
  As outlined in Sec.~\ref{theory}, noise and `overcorrection' can
  lead to a Gaussian pixel distribution that shows up as a linear run
  in the low column density range of the PDF. We therefore slightly
  iterated the correction value for the contamination value in order
  to avoid this effect and optimized the threshold where we start the
  lognormal fit.  The original Auriga PDF (see also Harvey et al.
  \cite{harvey2013}) shows a superposition of two PDFs where the first
  (low extinction) peak is compatible with the contamination level of
  the map and nearly disappears when a level of \dav\, = 0.8 is
  removed.  In the case of Auriga, the correction works very well
  because the contamination is most likely a rather homogeneous layer
  in front of or behind the Auriga cloud.  Such a superposition has
  already been observed in Pipe (Lombardi et al.
  \cite{lombardi2006}).
 
  The LOS contamination correction has several effects: (i) the PDF is
  broader (increase of $\langle \sigma_\eta \rangle$ from
  0.23$\pm$0.02 to 0.42$\pm$0.04), (ii) the slope becomes flatter, and
  (iii) the peak and deviation point of the PDF from lognormal to
  excess (\av(DP) from now on) shift to lower values.  In
  Sec.~\ref{theory}, we quantify these effects by an analytic model.
  The most dramatic change is observed for the PDFs of high-mass SF
  regions (Fig.~\ref{fig:pdf-gmc}).  The very narrow
  distribution\footnote{A low angular resolution (such as for
    extinction maps at a few arcmin resolution) also naturally results
    in narrower PDFs since the highest density structures are not
    resolved well. This effect can become important for very distant
    clouds. } ($\sigma_\eta$ = 0.27) for NGC3603 becomes much broader
  ($\sigma_\eta$ = 0.52), the PDF peak shifts from \av = 4.8 mag to
  1.6 mag, and \av(DP) from 8.1 mag to 4.9 mag.  The same applies for
  the Carina PDF with a change of $\sigma_\eta$ = 0.2 to 0.38 after
  correction\footnote{Our PDF is different from the one shown in
    Preibisch et al.  (2012) because their column density map was
    obtained from a fit using only the {\sl Herschel} wavelengths 70
    and 160 $\mu$m, which are not good tracers of {\sl \emph{cold}}
    gas.}.  Other {\sl Herschel} studies of high-mass SF regions, such
  as NGC6334 (Russeil et al.  \cite{russeil2013}) where a LOS
  contamination of \dav\ $\sim$2--3 mag was estimated, show the same
  narrow PDFs. {\sl \emph{It is thus essential to consider the
      contamination for a correct interpretation of PDFs for high-mass
      SF regions}}.
 
  In Table~\ref{table:summary} the PDF properties are listed, and they 
  reveal that the correction leads to an {\sl \emph{equalization}} of the 
  values for all clouds.  The PDF peak values now have a range of 1.4 
  mag to 2.6 mag instead of 2.3 mag to 4.8 mag, and \av(DP) changes to 
  3.5 mag to 5.5 mag with an average of 4.7$\pm$0.4 mag. (Original 
  values are 4.0 mag to 8.1 mag with an average of 6.8$\pm$1.0 mag.) 
  A value of \av(DP) between 4 and 5 is less than what was found in 
  other studies using extinction maps (\av(DP)=6 in Froebrich \& 
  Rowles 2010), but consistent with recent hydrodynamic models of 
  turbulent and self-gravitating gas (Ward et al. \cite{ward2014}). 
   
\begin{figure}   
   \centering   
   \includegraphics[width=8cm,angle=0]{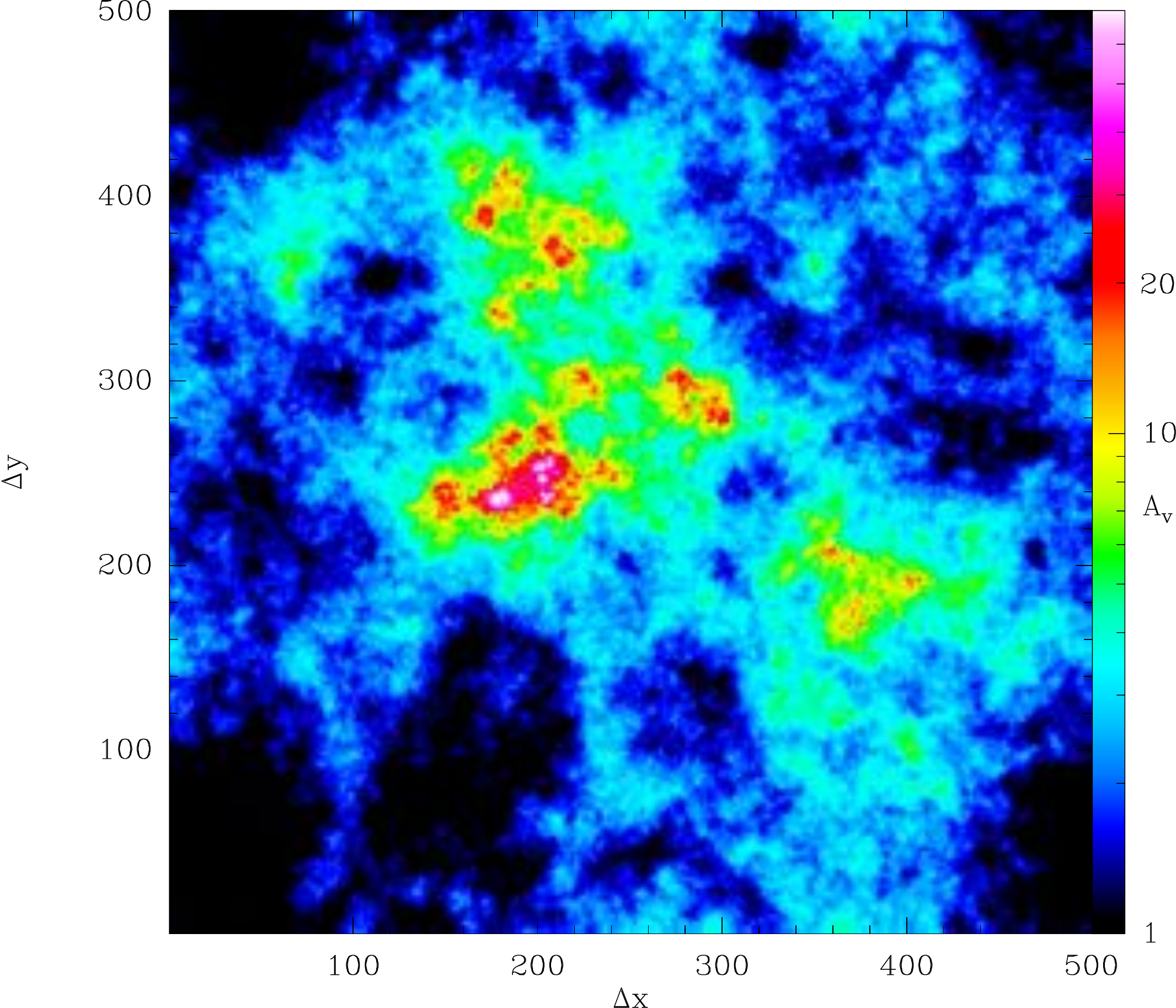}   
   \caption{Simulated map (500$\times$500 pixels) characterized by a 
     PDF with lognormal part and power-law tail, derived from a 
     fractional Brownian motion (fBm) map with a power spectral index 
     of 2.8.  The column densities are expressed in \av.} 
   \label{fbm-map}   
\end{figure}

\begin{figure}   
   \centering   
   \includegraphics[width=6cm,angle=90]{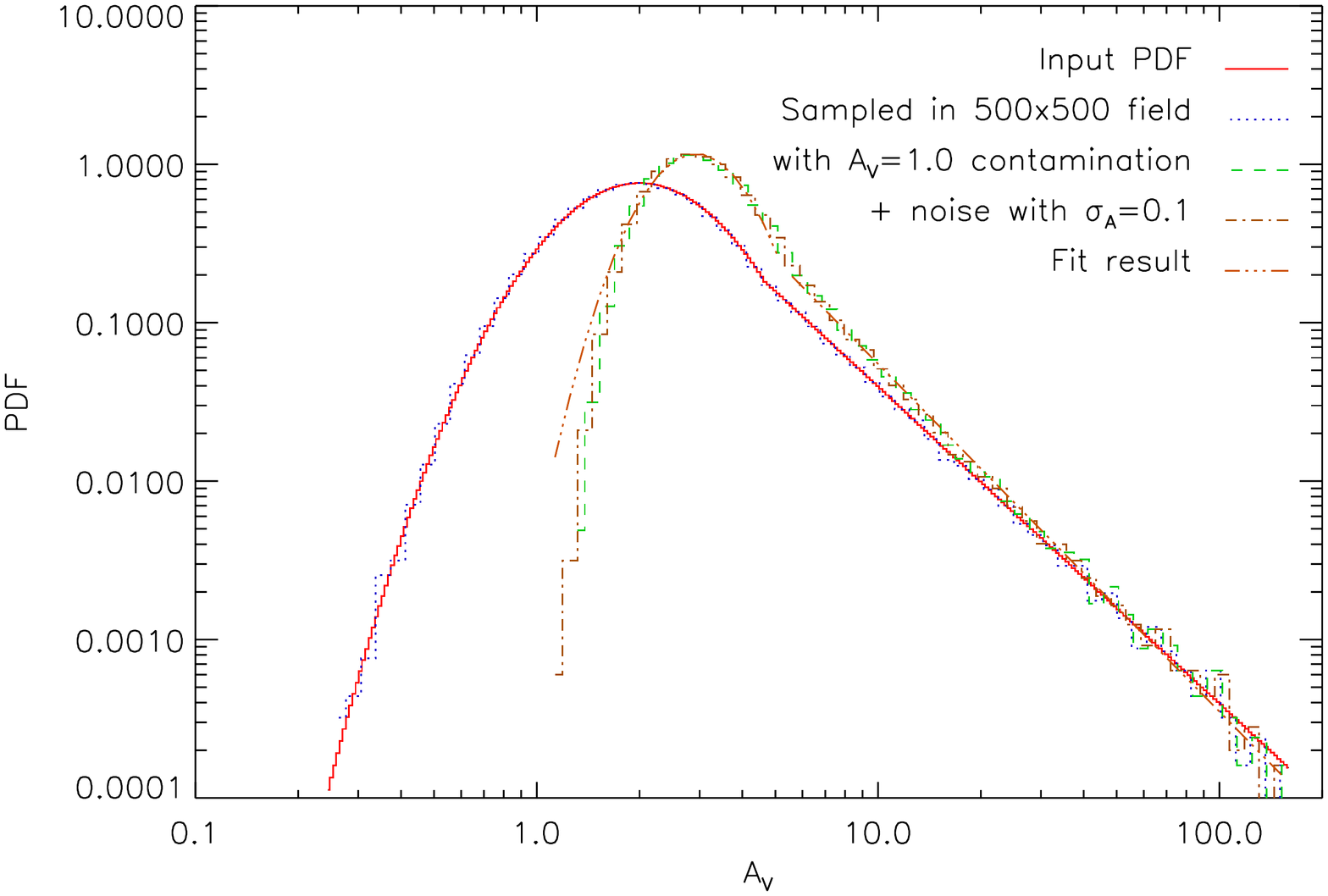}   
   \includegraphics[width=6cm,angle=90]{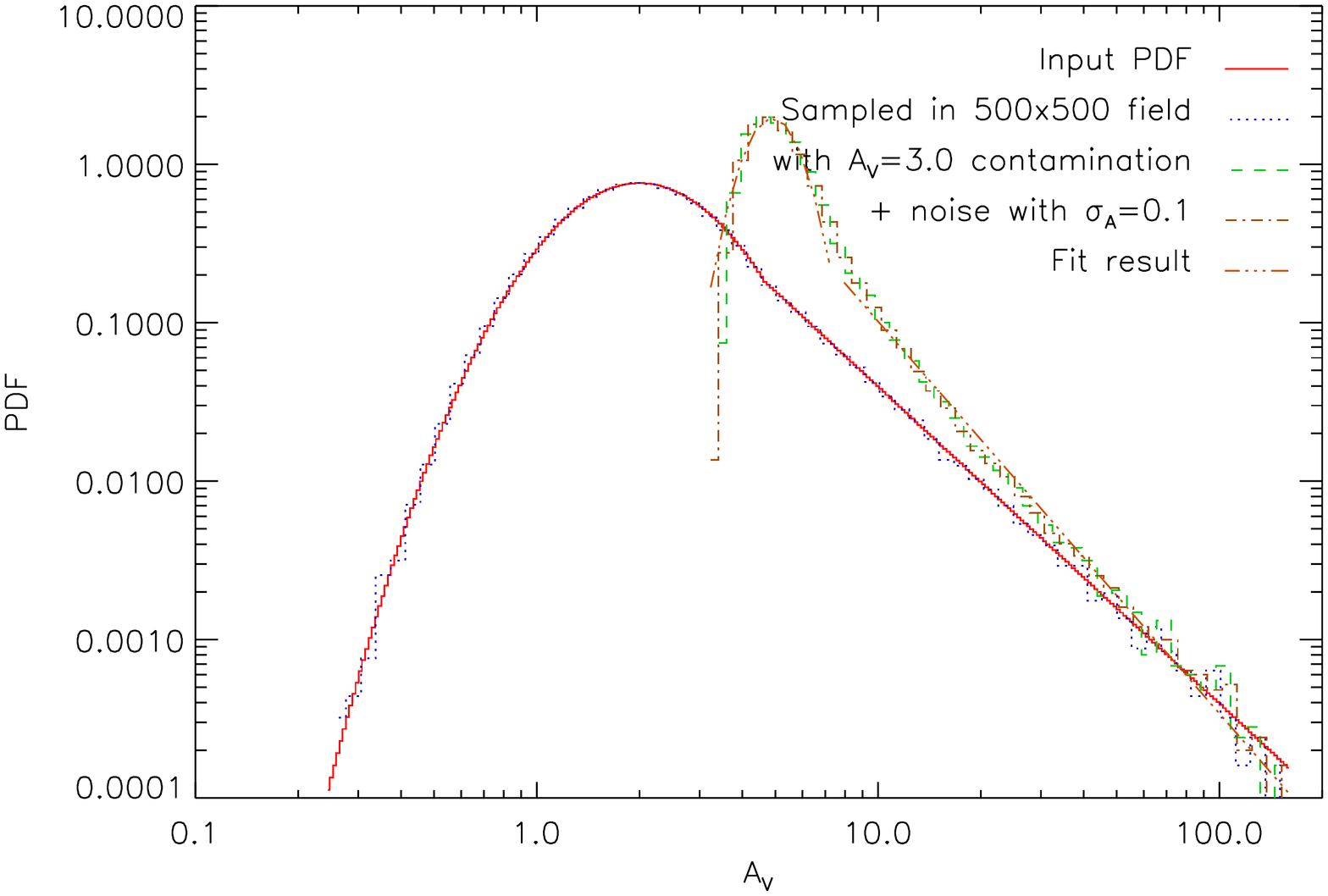}   
   \caption{Simulations of the PDF of an observed cloud with an  
     originally perfectly lognormal core and a power-law tail (solid line). The  
     representation by a finite-size map introduces some  
     uncertainties (dotted line).  The addition of an \dav\, = 1.0 (3)  
     contamination distorts the shape of the PDF and observational  
     noise adds a low-exctinction component. The deviation point  
     \av(DP) shifts by $\sim$4 mag for a contamination  
     of \dav\, = 1.0 (3).  The fit of the resulting PDF by a  
     lognormal core and a power-law tail does not recover the initial  
     parameters.}  
   \label{fig_shift1_0}   
\end{figure}

\section{Simulations}   \label{theory}    
  
\subsection{Method} 
 
We modeled the effect of LOS contamination in a numerical simulation,
providing `ideal' PDFs consisting of a lognormal part and a power-law
tail. Maps were generated with 500$\times$500 pixels matching the
typical observational map size and grid investigated in
Sect.~\ref{coldens}.  To simulate maps that combine the given
probability distribution of pixel values with spatial correlations
among the pixels that are characteristic of observed maps (needed in
6.3), we started from fractional Brownian motion (fBm) fractal maps
(Stutzki et al. \cite{stutzki1998}) with a power spectral index of
2.8, matching the typically observed spatial scaling behavior. Because
the fBm maps are characterized by a normal distribution of values, we
obtained the desired PDF in a second step by shifting each density by
the value given by the ratio of the inverse integrals of the original
and the desired PDFs. The result is visualized in Fig.~\ref{fbm-map}.
Larger and smaller fields were tested to verify the numerical accuracy
of the method.
 
The generated field was then ``contaminated'' by adding a constant
level to all map values and finally ``observed'' including Gaussian
white noise typical of the {\sl Herschel} observations.  The overall
model is characterized by six parameters: the width and the center
\av\ of the lognormal PDF contribution, the exponent of the power-law
tail, the deviation point characterizing the transition from the
lognormal to the power-law PDF, the additive contamination \dav\, and
the standard deviation of the observational noise. To analyze the
impact of the contamination we fixed all other parameters to lie
within a range obtained for the observations in Sect.~\ref{coldens},
i.e., a center (peak) at \av = 2.0 mag, $\sigma_\eta=0.5$, $s=-2.0$,
\av(DP) = 4.3 mag, and a noise rms $\sigma_{A_V}=0.1$.
 
\subsection{Effect of LOS-contamination on  PDF properties }  
  
In Fig.~\ref{fig_shift1_0} it becomes obvious that the fit of the 
resulting measurable PDFs provides parameters that clearly deviate 
from the original input, in particular for the \dav=3.0 case.  One 
observes that the lognormal part of the PDF is strongly 
compressed, consistent with what we observe for the NGC3603 cloud 
(Fig.~\ref{fig:pdf-gmc}). The statistical sampling of that part 
becomes very rough.  The value for \av(DP) increases, and the slope of 
the power-law tail becomes steeper than $s=-2.0$ that characterized 
the original cloud.\footnote{The offset correction 
  mathematically results in a modified function without a power-law 
  tail. However, the corrections are small enough, so that fitting a 
  power-law function and inferring a slope from this fitting function 
  is still a reasonable procedure.} 
  
\begin{figure}   
   \centering   
   (a)\includegraphics[width=5.5cm,angle=90]{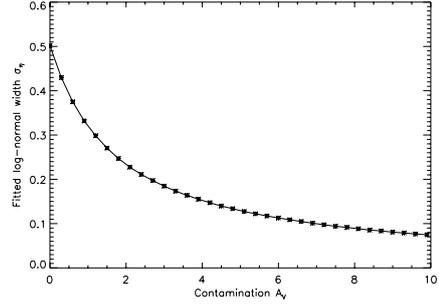}\\   
   (b) \includegraphics[width=5.5cm,angle=90]{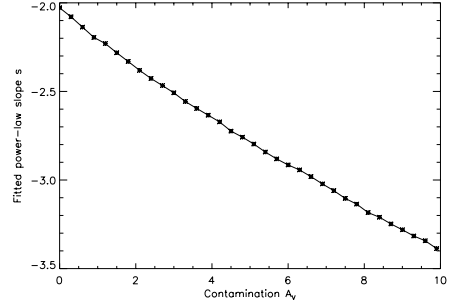}\\   
   \caption{Dependence of the fitted width of the lognormal   
   core of the PDF {\it (a)} and the fitted slope of the power-law   
   tail {\it (b)} as a function of the foreground (and/or background) contamination   
   for the standard parameters of the molecular cloud PDFs.}   
   \label{fig_scan}   
\end{figure}   
   
To quantify this effect systematically, we show in Fig.~\ref{fig_scan} 
the change in the width of the PDF and the slope of the power-law tail 
when measured as a function of LOS contamination, always starting from 
the standard parameters of the underlying cloud with $\sigma_\eta=0.5$ 
and $s=-2.0$. The small irregularities in the curve result from the 
discrete binning of the randomly sampled PDF. We find a dramatic 
effect in both parameters. For a contamination of \dav=2, 
$\sigma_\eta$ is already reduced by more than a factor two, and the 
slope of the power-law tail has steepened from $-2.0$ to $-2.4$. For 
an \dav$\approx 10$, characterizing distant massive regions or 
infrared dark clouds, $\sigma_\eta$ has decreased by more than a 
factor of five and the power-law tail has a slope of $-3.4$. 
  
\begin{figure}   
   \centering   
    \includegraphics[width=6cm,angle=90]{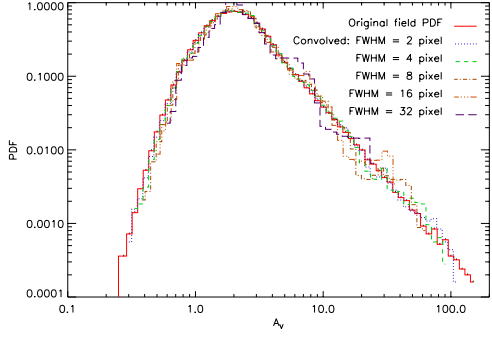}   
    \caption{Influence of finite beam size on the PDF shape.   
      Different beam size (expressed as pixel size of 2, 4, 8, 16, and 32)   
      were chosen. The convolution has a minor effect for the 
      high-column density tail of the PDF and no influence on the 
      lower-density lognormal distribution. } 
   \label{fig_reso}   
\end{figure}   
 
\subsection{Effect  of finite resolution on the PDF}  
 
Figure~\ref{fig_reso} shows the impact of a finite angular resolution
on the PDF shape, which is implemented here as a convolution with a
Gaussian beam of varying FWHM. We find the impact of a reduced
resolution at very low column densities and at column densities above
\av = 15.  When reducing the resolution, the measured peak column
densities drop.  However, one clearly sees that a finite resolution
only detracts from the highest peaks in the high-density PDF tail, but
it does not change any of the properties of the PDF (peak, width,
deviation point, slope) that we discuss here. However, lower
resolution can lead to bumps and distortions in the general power-law
tail.  In real maps with observational noise, the change at low
densities would not be visible because the low column density part is
more strongly affected by noise.
 
\begin{figure}   
   \centering   
    \includegraphics[width=6cm,angle=90]{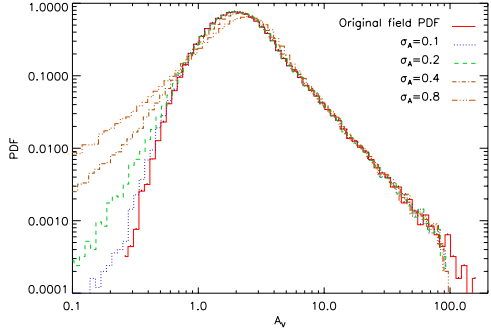}   
    \caption{Simulations of reconstructed PDFs observed with different 
      amplitudes of noise $\sigma_{\rm A}$, expressed in terms of 
      visual extinction. The reconstruction does not depend on the 
      absolute value of the foreground/background contamination.} 
   \label{fig_noise}   
\end{figure}   
 
\subsection{Effect of noise and uncertainty on the correction}  
  
The addition of observational noise adds a contribution to the
structure that is hardly visible in the contaminated PDF, but that may
reappear as amplified after the correction of the contamination. The
noise is assumed to have an approximately Gaussian distribution
providing a floor of small fluctuations with zero average. In the
logarithmic binning, the core of this Gaussian gives a linear
contribution to the PDF for low densities, i.e. below the lognormal
part of the PDF of the observed structure. This effect is simulated in
Fig.~\ref{fig_noise} where we varied the observational noise from
Fig.~\ref{fig_shift1_0} and applied the contamination correction to
reconstruct the PDF of the original structure.
  
We find a clear excess at low column densities, which turns close to 
the expected linear behavior for $\sigma_{A_{\rm V}}=0.8$.  Starting 
from $\sigma_{A_{\rm V}}=0.4,$ we also find a slight, but noticeable, 
shift in the PDF peak to higher \av. For actually observed data, this 
should be taken into account.  The effect is independent of the 
contamination that is added and subtracted in the PDF transformation. 
  
The observed PDFs in Sect.~\ref{coldens} show the same kind of
low-column density excess as these simulations; in fact, the example
of the Maddalena cloud matches the simulation for $\sigma_{A_{\rm
    V}}=0.4$ exactly. However, the pure observational noise in the
column density maps is much lower.  This can be explained by
additional small scale uncertainties and fluctuations in $\sigma_{\rm
  A}$. They behave similarly to noise, but are not observational
noise.  They may represent fluctuations in the overall cloud
contamination, either variations in the foreground screen or small
background contaminating clouds (see also discussion in Alves et al.
(2014) for correlated pattern of noise in extinction maps).  As long
as the fluctuations are relatively small, their impact on the fit of
the main lognormal part of the PDF is negligible.
  
\begin{figure}   
   \centering   
   \includegraphics[width=6cm,angle=90]{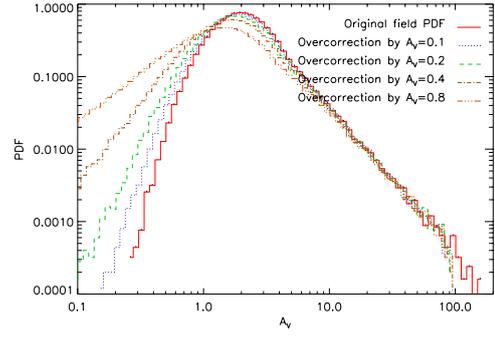}   
    \caption{Simulations of a PDF with negligible noise  
    ($\sigma_{\rm A}$ = 0.01) when applying values that are too high  
    for the contamination correction \dav. The reconstruction   
    does not depend on the absolute value of the foreground/background   
    contamination but only on the difference between actual   
    contamination and subtracted contamination.}  
   \label{over}   
\end{figure}   
  
Very low column densities, giving rise to a linear contribution to the
PDF for low \av{}, can also stem from an ``overcorrection'' of the
contamination, however, shifting part of the real cloud structure to
column densities around zero. Therefore, using too high a value for
the contamination correction by \dav, leads to a similar effect as
increased noise. Figure~\ref{over} shows this for an example
calculated with negligible noise ($\sigma_{\rm A}$ = 0.01) for
different levels of \dav. We find a similar linear distribution for
the low column-density pixels that becomes more important for
increasing \dav. In this case, the peak of the PDF shifts in the other
direction, compared to the addition of noise/fluctuations, i.e., to
lower column densities.
  
\begin{figure}   
   \centering   
   \includegraphics[width=5.5cm,angle=90]{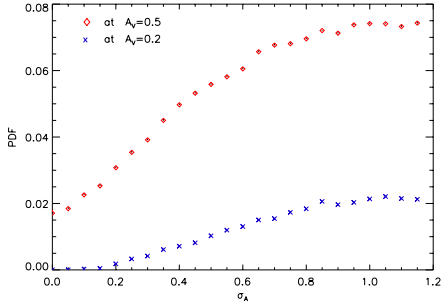}   
    \caption{Amplitude of the low column density in  
    the PDF as a function of the added level of noise or   
    small-scale fluctuations, corresponding to Fig.~\ref{fig_noise}  
    for two points in the low-column density wing. The scatter  
    in the points results from the adaptive binning in the PDF  
    computation.}  
   \label{fig_errorestimate}   
\end{figure}

\begin{figure}   
   \centering   
   \includegraphics[width=8cm,angle=90]{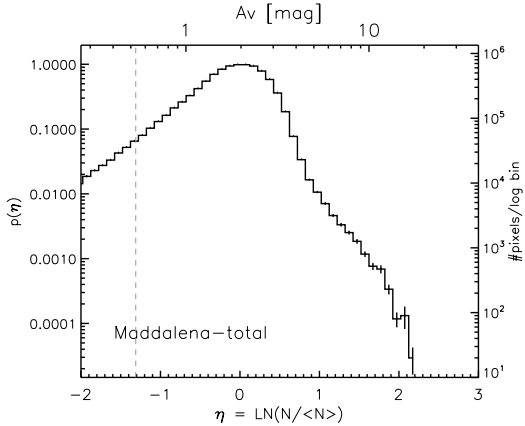}   
    \caption{PDF of Maddalena including the very low column density range down to  
\av $\sim$0.2 mag. The dashed line indicates the \av = 0.5 mag level used to derive the amplitude  
    of the PDF (in this case $\approx$0.06) to be compared to Fig.~10 in order to determine the  
uncertainty of the correction (in this case $\approx$0.5-0.55).}  
   \label{error-mad}   
\end{figure}   
  
When evaluating the low-column-density excess, we find that an 
overcorrection always has a stronger impact than noise or fluctuations 
of the same amplitude. Therefore we can analyze the low-density excess 
in the observations to get an upper limit of the uncertainty in the 
contamination correction when comparing the excess with the noise 
impact. This uncertainty then represents either fluctuations in the 
contamination or an absolute error in the contamination correction (in 
which case it would be somewhat overestimated). This is done through 
the simulation shown in Fig.~\ref{fig_errorestimate}. It gives the 
amplitude of the PDF at $A_{\rm V}=0.2$ and $A_{\rm V}=0.5$ as a 
function of the amplitude of the fluctuations; i.e., this quantifies 
the low-density excess in Fig.~\ref{fig_noise} for all possible 
noise/fluctuation levels.  For all observations, we used this 
approach to quantify the uncertainty of the contamination correction, 
then excluding the low-density part from the fit of the lognormal part 
of the PDF of the actual cloud structure (see Sect.~\ref{coldens}). 
Figure~\ref{error-mad} illustrates an example of this procedure where 
we chose the `worst case' scenario for Maddalena.  The PDF is plotted 
over a wide column density range, and the \av = 0.5 mag level 
corresponds to an amplitude of the PDF of $\approx$0.06. Using Fig.~7 
as a look-up table indicates that for this amplitude, the uncertainty 
of the correction corresponds to approximately 0.5 magnitudes. 
  
In this way, we not only can estimate the contamination of the cloud
extinction by foreground or background material, but also the
uncertainty of the contamination. The uncertainty can be due to random
fluctuations or a systematic error. For any measured low-density
excess, the systematic error is always smaller than the random error,
so that we provide an upper limit to the total uncertainty by giving
only the magnitude of the possible fluctuations.
 
Summarizing, these findings clearly demand a correction of the
contamination when analyzing observed column-density or extinction
maps.  As a first easy approach, the contamination can be removed from
the maps before the analysis, or Fig.~\ref{fig_scan} can be used as a
lookup table to obtain the original parameters from the measured width
and power-law slope.  In the next section, we discuss the physical
implications (mean column density, column density structure, and
cumulative mass function) of the corrected {\sl Herschel} column
density maps.
  
\begin{figure*}[!htpb]    
\centering    
\includegraphics [width=7cm, angle={0}]{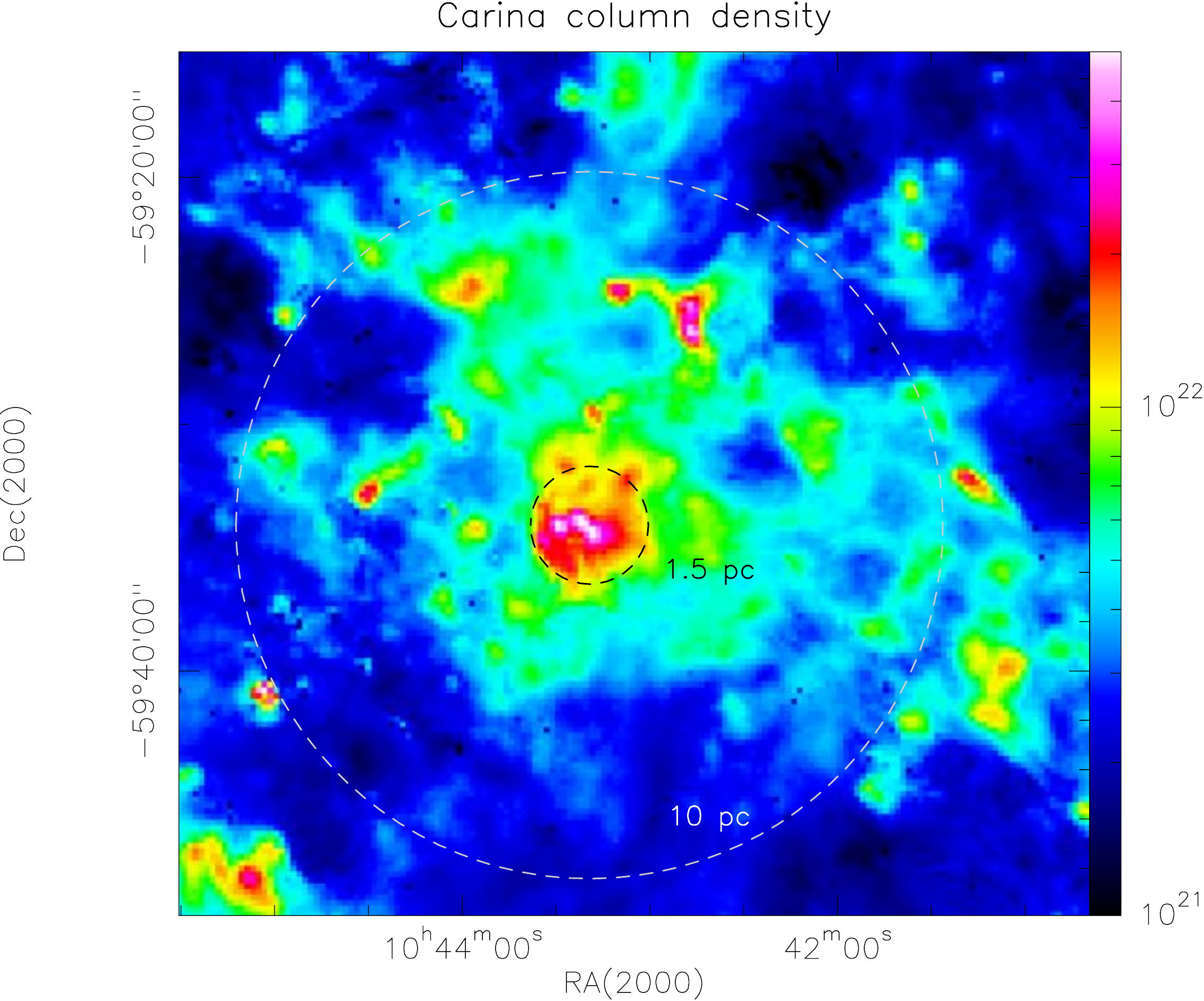}    
\hspace{0.1cm}\includegraphics [width=7cm, angle={0}]{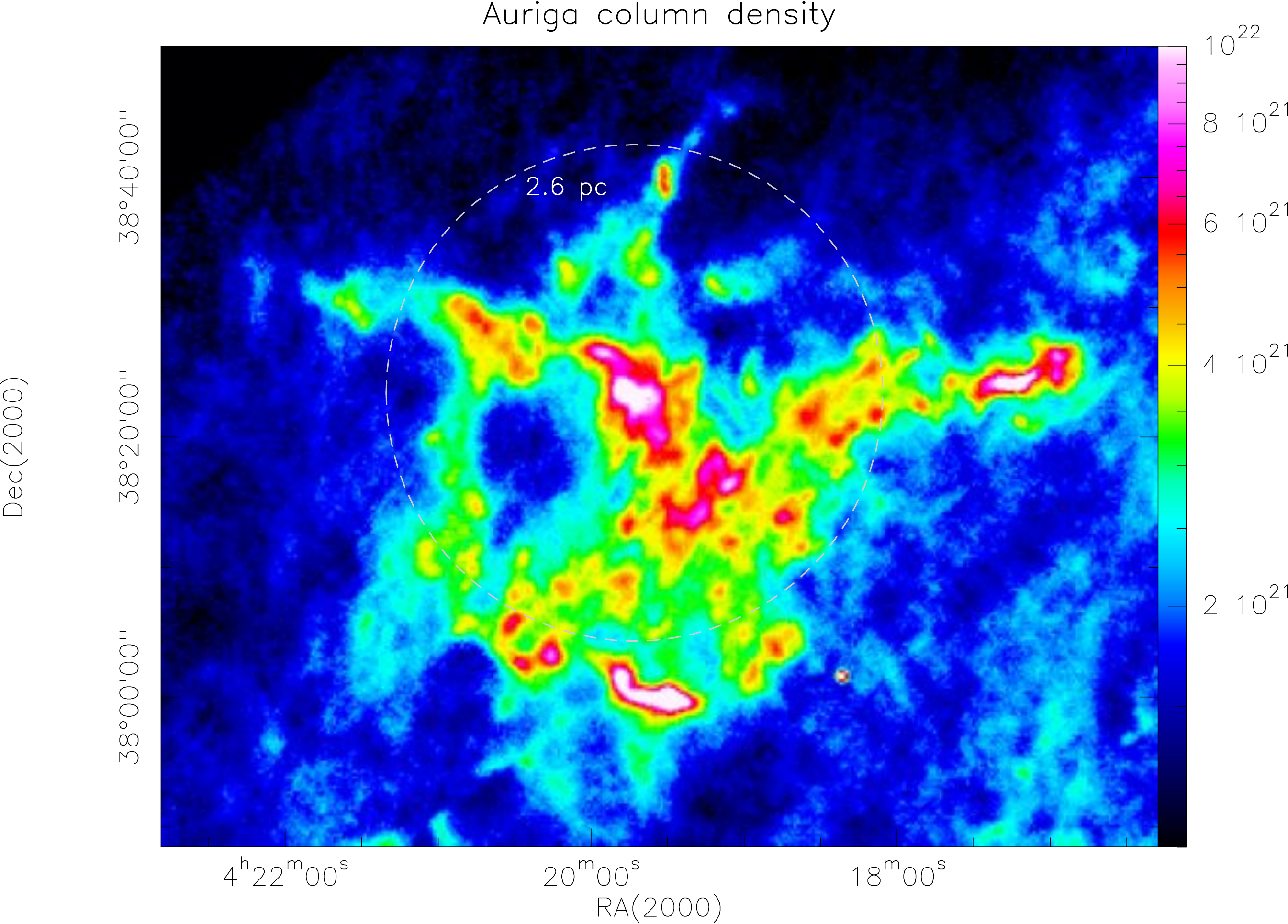}    
\includegraphics [width=7cm, angle={0}]{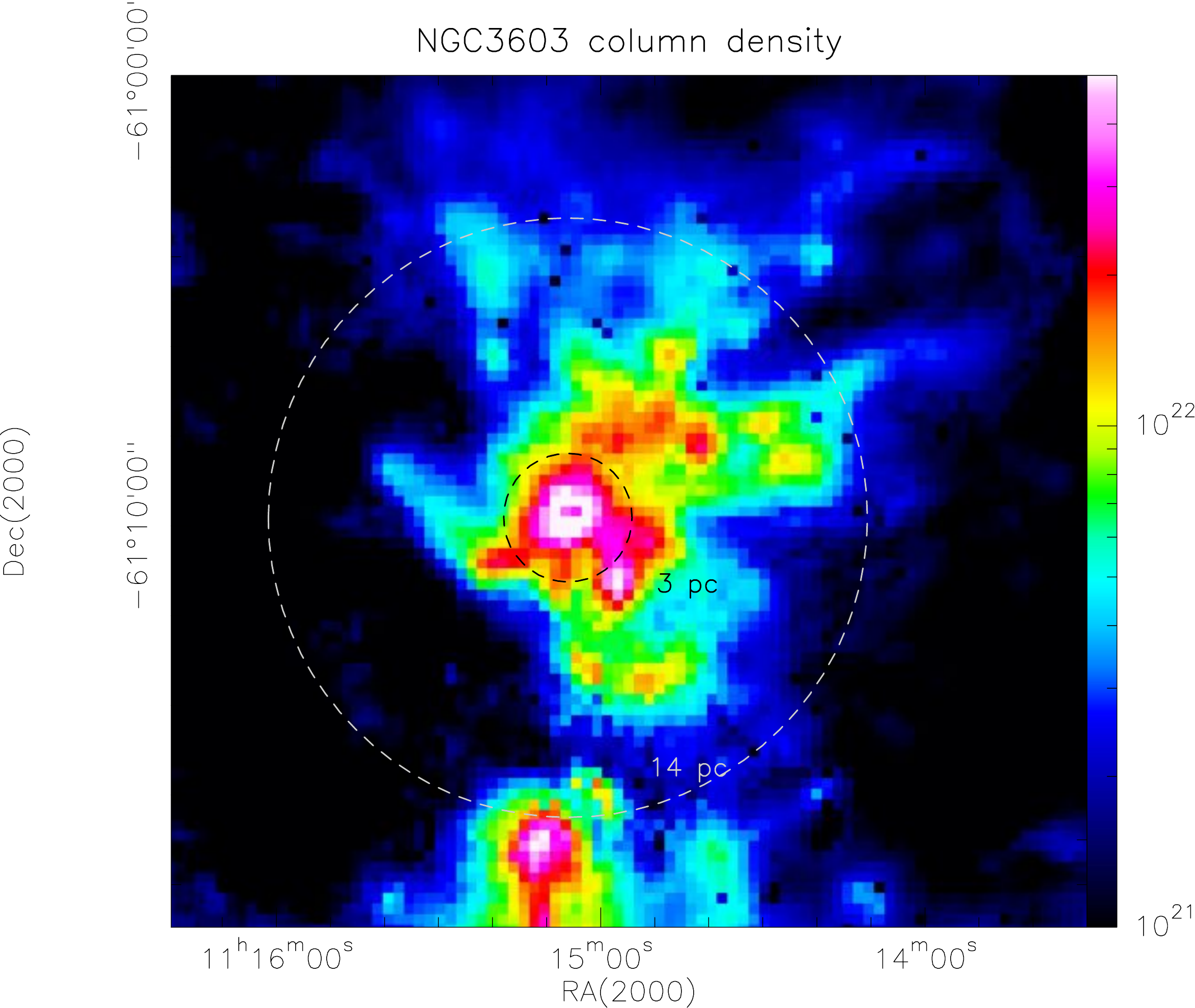}    
\hspace{0.1cm} 
\includegraphics [width=7cm, angle={0}]{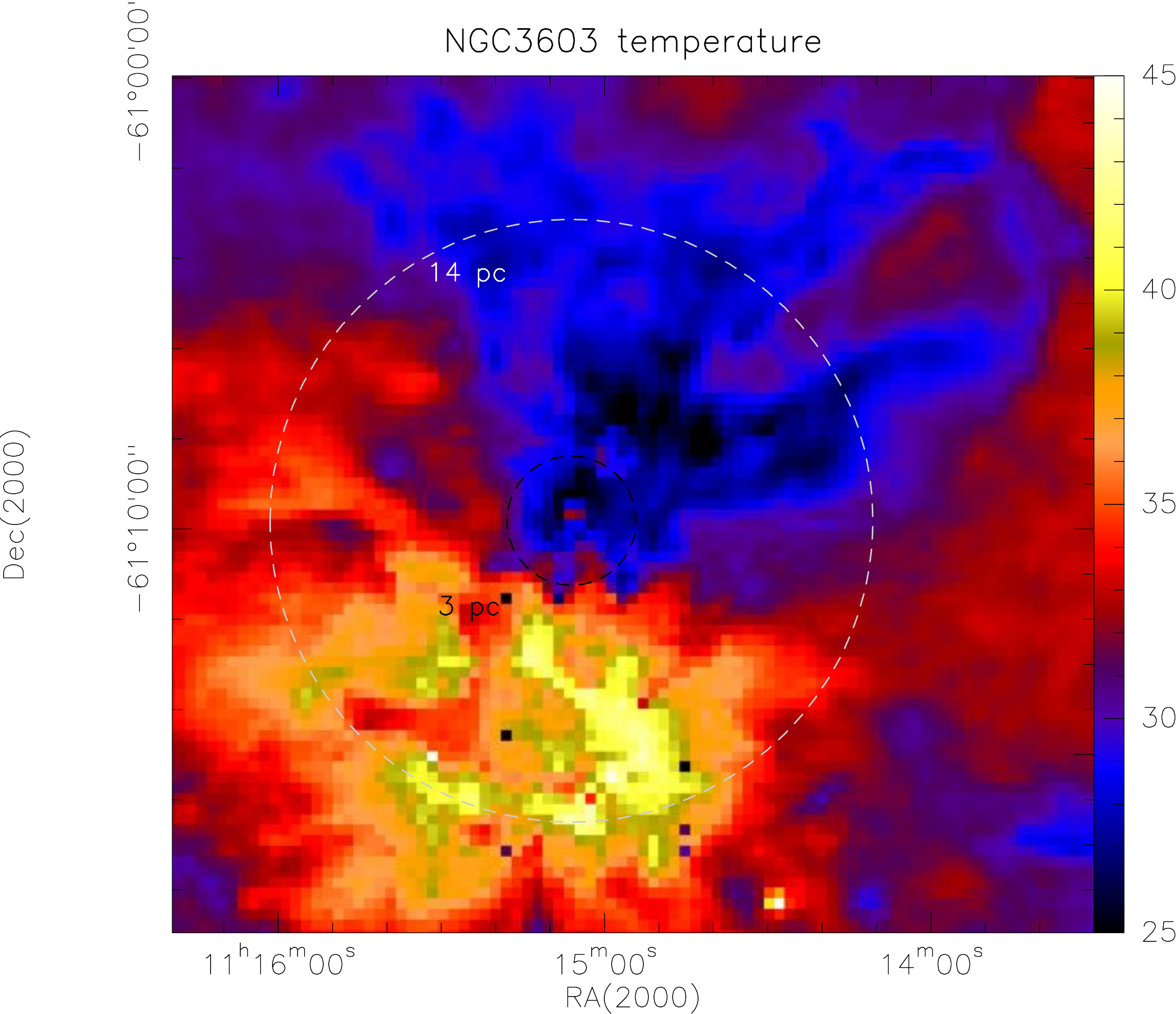}    
\caption[] {Zoom into the column density maps of NGC3603, Carina, and 
  Auriga where we determined the column density profiles shown in 
  Fig.~\ref{profile}. The lower right panel shows a temperature map of 
  NGC3603.  The white, dashed circle includes all the pixels we included 
  for the profile and the black, dashed circles for NGC3603 and Carina 
  outline the area where we observe a possible slope change in the 
  column density profile.} 
\label{circles}    
\end{figure*}    
  
\section{Discussion }   
  
\subsection{Do all molecular clouds have the same mean column density  ?}   
  
The third Larson law (Larson \cite{larson1981}) is an empirical
relation between the volume density $\rho$ of a cloud and its size $r$
with $\rho \propto r^{-1.1}$, and this implies that all molecular
clouds have approximately constant mean column densities. However,
there is a long-lasting controversy about its validity in theory
(e.g., Ballesteros-Paredes \& Mac Low \cite{ball2002}) and
observations (e.g., Schneider \& Brooks \cite{schneider2004}, Heyer et
al.  \cite{heyer2009}). Recently, a study of Lombardi et al.
(\cite{lombardi2010}), using extinction maps of five molecular clouds,
has concluded that all clouds in their sample have identical average
column densities above a given extinction threshold.  However, their
investigation does not include very massive and dense clouds. (The
only high-mass SF region in their sample is Orion A with a column
density below 10$^{23}$ cm$^{-2}$, see, e.g. Polychroni et al.
\cite{poly2013}.)
  
We show that a correction of column density maps for the effect of LOS
contamination leads to a typical value of 1.5--3 $\times$10$^{21}$
cm$^{-2}$ for the average column density for the individual clouds
with a total average for all clouds of $\langle N(H_2) \rangle =
(2.4\pm0.4) \times$10$^{21}$ cm$^{-2}$.  This range is fairly narrow,
taking the whole dynamic range of column densities of molecular clouds
into account across two magnitudes (a few 10$^{21}$ cm$^{-2}$ up to a
few 10$^{23}$cm$^{-2}$).  There is a tendency for the high-mass SF
clouds in our sample, NGC3603 and Carina, to have slightly higher
average column densities and surface densities compared to the
low-mass SF clouds Auriga and Maddalena, but this difference in a
sample of only four clouds is not yet significant and needs further
studies. In any case, similar mean column densities for all cloud
types do not imply that the global column density structure of all
molecular cloud types is the same, as we discuss in the next section.
 
\subsection{Radial column density profiles} \label{cd-profiles} 
  
\begin{figure*}[!htpb]    
\centering    
\includegraphics [width=12cm, angle={0}]{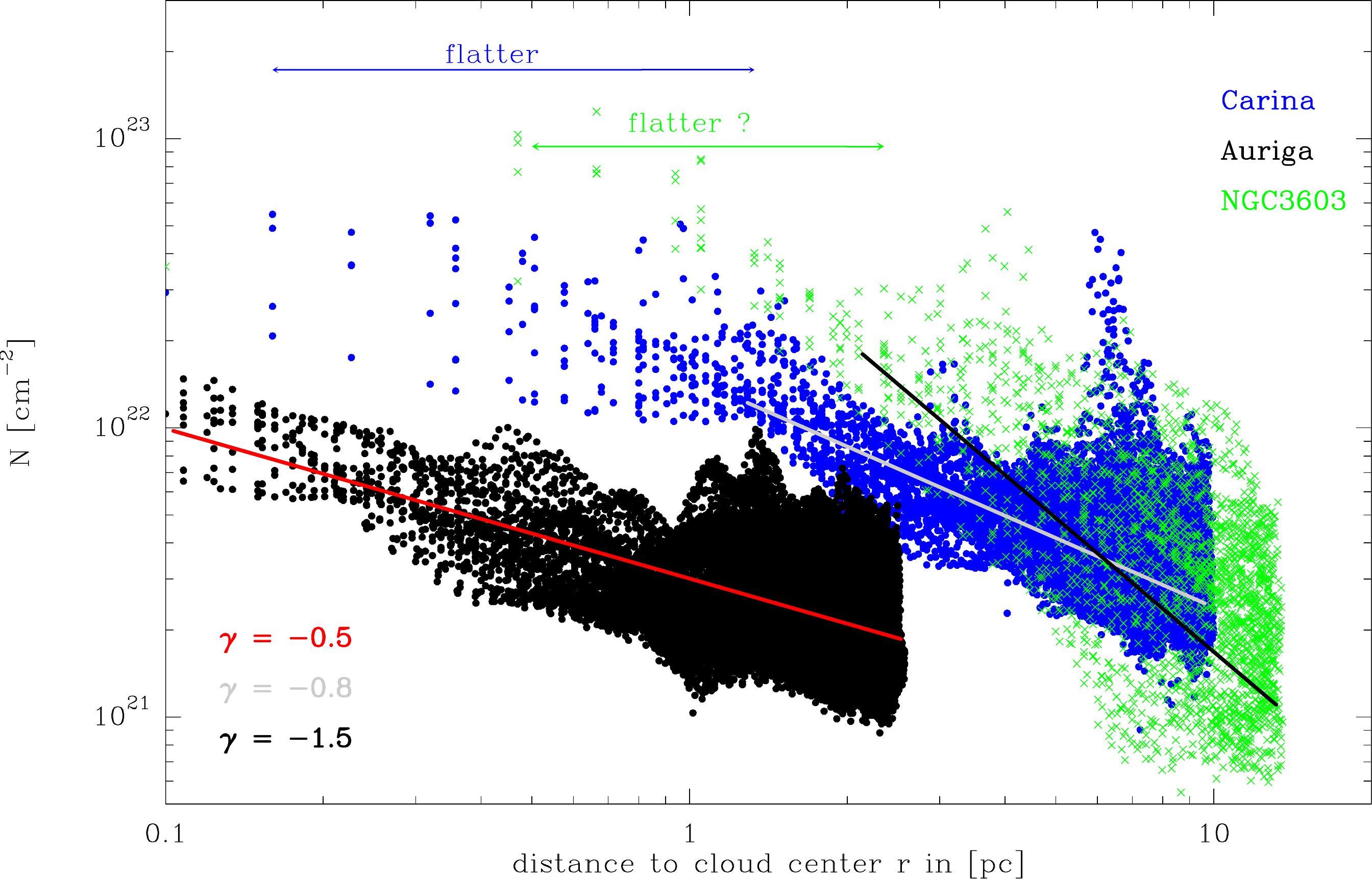}    
\caption[] {Column density profiles for NGC3603, Carina, and Auriga. 
  The angular resolution is 0.03, 0.16, and 0.47 pc for Auriga, 
  Carina, and NGC3603, respectively.  The result of linear regression 
  fits are plotted as a red (Auriga), gray (Carina), and black 
  (NGC3603) line, the slopes are given in the panel (the error on the 
  slopes is $\sim$0.1).} 
\label{profile}    
\end{figure*}    
 
Though the average column density in various cloud types shows no
strong distinction, there is still a small difference between the
low-mass SF regions Maddalena and Auriga ($\langle N \rangle$ =
1.8$\times$10$^{21}$ cm$^{-2}$) and the high-mass SF regions Carina
and NGC3603 ($\langle N \rangle$ = 3.0$\times$10$^{21}$ cm$^{-2}$).
Is this variation possibly because the clouds have a different column
density {\sl \emph{structure}} across the cloud, with a common
low-density regime but accumulation of denser gas in smaller volumes ?
 
To test this, we determined radial column density profiles for three
of our clouds in the sample.  Figure~\ref{circles} shows NGC3603 and
Carina, which are centrally condensed enough to enable such a study,
while the Auriga cloud is very filamentary so that we selected only
one prominent clump (in the center of the filament, see
Fig.~\ref{cds}) with a somewhat circular shape. We excluded Maddalena
because there is no such clump. We then determined the column density
at each point on the maps depending on the distance to the center and
plotted the resulting profiles in Fig.~\ref{profile}. The clouds are
`cropped' because we need to focus on these spherical central regions.
 
At first glance, all profiles show no clear jump in column density at
a certain threshold level in column density or radius. Because all
clouds have significant substructure, some denser clumps and filaments
protrude as peaks in the column density profile (e.g., at 6 pc for
Carina). Ignoring these peaks, the profiles show a rather smooth
slope, though there is a tendency to a shallower slope for Carina and
NGC3603 toward the cloud center, i.e. lower radii $r$.  The profiles
were not deconvolved from the beam so that the result for NGC3603
(resolution of 0.47 pc) is less reliable.  Tremblin et al.  (2014) and
Didelon et al.  (in prep.) have observed such a flattening of the
slope toward the cloud center in M16 and Mon R2, which are both
regions with closely associated \hii\ regions.  Most important, the
slopes are not the same for the three clouds.  A fit of $N \propto
r^\gamma$ over the whole range of column density and radii (starting
at the resolution limit and ignoring a possible flattening toward the
cloud center) gives values of $\gamma$ = --0.5$\pm$0.08,
--0.8$\pm$0.1, and --1.5$\pm$0.15 for Auriga, Carina, and NGC3603.
The steeper slope of NGC3603 already becomes obvious purely by visual
inspection. Assuming a centrally condensed, spherical density
distribution, this corresonds to density profiles $\rho(r) \sim N \,
r^{-1} \sim r^{\gamma -1} \sim r^{-\alpha}$ with $\alpha$ = 1.5, 1.8,
and 2.5.  These values correspond reasonably well to the ones we also
derived from fitting the power-law tail of the PDF (1.78, 1.75, 2.5)
where we assumed an equivalent density profile (see the studies of the
link between PDF and self-gravity by Klessen (\cite{klessen2000}),
Kritsuk et al.  (\cite{kritsuk2011}), Federrath \& Klessen (2013),
Girichidis et al.  (\cite{giri2014})).  In any case, we do not expect
a one-to-one correspondence because we do not sample exactly the same
areas.
 
The collapse of an isothermal sphere has been studied for a long time
(Larson \cite{larson1969}, Penston \cite{penston1969}, Shu
\cite{shu1977}, Whitworth \& Summers \cite{whitworth1985}), and though
all models start with different initial conditions, they arrive at the
same $\alpha$ = 2 for early stages and $\alpha$ = 1.5 after a
singularity formed at the center of the sphere.  Our slope
determinations from the column density profiles and the power-law
tails now clearly point toward a scenario in which gravitational
collapse is the dominating physical process. We cannot judge whether
the cloud as a whole is collapsing or only the individual clumps,
cores, and filaments inside the cloud. However, from a theoretical
point of view, this does not make any difference.  Girichidis et al.
(2014) state that {\sl ``If each density parcel collapses on its
  free-fall timescale then the evolution of the overall PDF is
  independent of the number of fragements and the details of the
  fragmentation process.''}
 
Another argument for the importance of {\sl self-gravity} comes from 
the Jeans instability. The number of Jeans masses\footnote{$M_J 
  \propto T^{3/2} \, \rho^{-1/2}$ with temperature T = 20 K as a 
  typical average value, and density $\rho = M/V$, assuming uniform 
  density, and volume $V = (M/\Sigma)^{3/2}$ with the mass $M$ and 
  surface density $\Sigma$ of the cloud from 
  Table~\ref{table:summary}} $N_J = M/M_J$ of our clouds in the sample 
is 2040 for NGC3603, 1350 for Carina, 1100 for Maddalena, and 110 for 
Auriga.  Even under the simplified assumption of uniform density, 
which gives a lower limit on the number of Jeans masses, all clouds 
are orders of magnitude more massive than the average Jeans mass. 
Moreover, from the structural analysis we see that all of the objects are 
not at uniform density, but are highly structured. The Jeans mass in the 
regions of interest thus drops, which gives even higher numbers. For a 
density contrast of 100, which is certainly not an unrealistic 
assumption, the above values increase by an order of magnitude, 
implying that we definitely should expect signs of self-gravity. 
 
It should be noted that for NGC3603, we find an $\alpha$ that is
larger than two that does not correspond to free-fall alone.  A
possible explanation is radiative feedback.  Tremblin et al. (2014)
showed that the power-law tail of the PDF becomes flatter going from
the cloud center toward the interaction zone between an HII region and
the cloud. This implies that compression of gas takes place and that
self-gravity then takes over in the densest regions to form cores and
finally stars. The flat slope of the PDF for NGC3603 is thus probably
also a consequence of radiative feedback. In Fig.~\ref{circles} we
show a temperature map of NGC3603 where the \hii\ region becomes
apparent as a high temperature region southeast of the densest and
coldest parts of the molecular cloud.
 
Coming back to the question of whether all molecular cloud types have
the same column density structure, the answer seems to be no.  The
structure is the same at low and moderate column densities up to the
start of the PDF tail, but from the column density profiles and the
flatter power-law tails in the PDF seen for the massive SF regions
NGC3603 and Carina, we find an assembly of denser gas in smaller cloud
volumes compared to low-mass SF regions. Studying the distribution of
mass at different \av\ scales (see next section) illustrates this
behavior.

\subsection{Distribution of masses} \label{masses}  
 
\begin{figure}[!htpb]    
\centering    
\includegraphics [width=8cm, angle={0}]{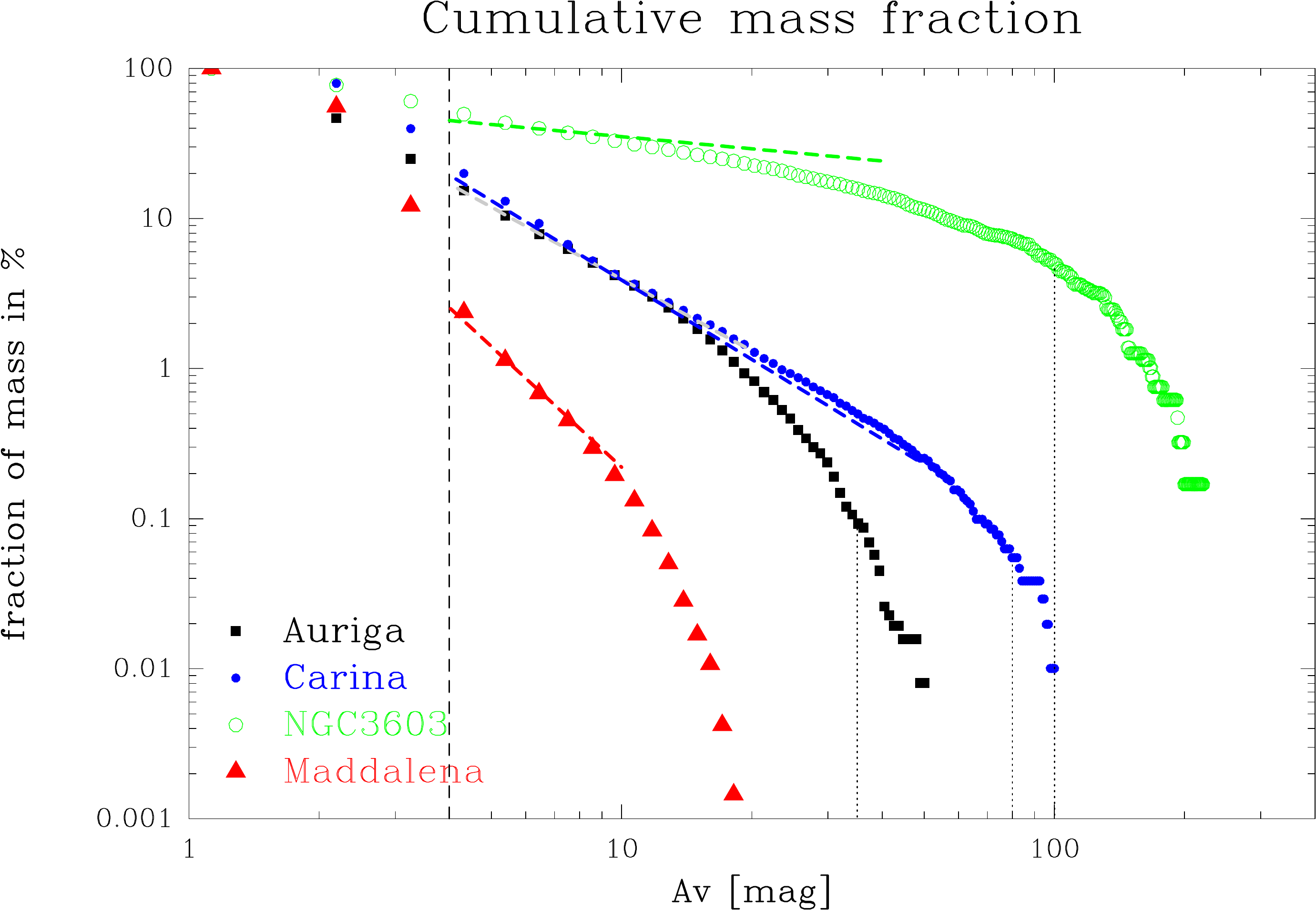}    
\caption[] {Log-log display of the mass fraction in percent above the 
  \av\,-level indicated in the x-axis for the four clouds in our 
  sample. The curves with the respective slopes from the PDF are 
  overplotted as dashed lines.  The short dotted vertical lines 
  indicate the conservative \av\,-level above the pixel number becomes 
  low and thus less reliable.  The long dashed line outlines \av\ = 
  4.} 
\label{mass-log}    
\end{figure}    
 
\begin{figure}[!htpb]    
\centering    
\includegraphics [width=8cm, angle={0}]{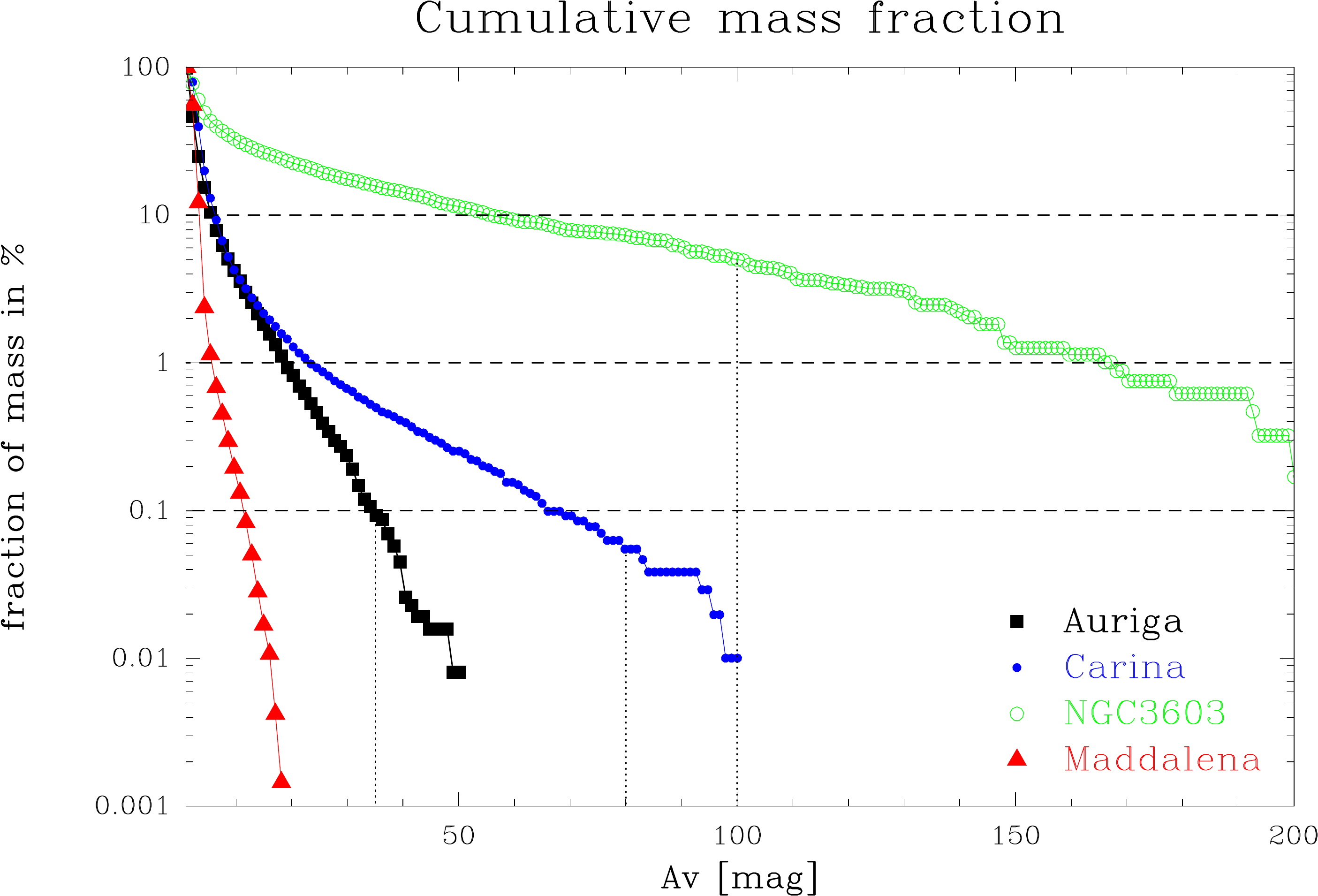}    
\caption[] {Log-lin display of the mass fraction in percent above the 
  \av\,-level indicated in the x-axis for the four clouds in our 
  sample. The short dotted vertical lines indicate the conservative 
  \av\,-level above the pixel number becomes low and thus less 
  reliable. } 
\label{mass-lin}    
\end{figure}    
 
Figures~\ref{mass-log} and \ref{mass-lin} show the fraction of mass in
percent at different \av\,-levels, i.e. column density thresholds (in
log-log space and log-lin space to be consistent with Lada et al.
(2010) and Kainulainen et al. (\cite{kai2013})).  In general, the
cumulative mass fraction is dominated by the column densities (masses)
making up the power-law tail of the PDF.  Figure~\ref{mass-log} shows
that the curves follow a power-law with an exponent of $s+1$,
resulting from the direct integration of the PDF to obtain the
cumulative mass function, until a fall-off where the PDF becomes noisy
due to a lower statistics (which is also true for the cumulative mass
functions, see Figs.~\ref{mass-log} and \ref{mass-lin}).  In principle
the cumulative mass function is derived from the PDF as
\begin{eqnarray*} 
M(A_V) \propto \int_0^{A_V} {d N(A_V') \over d A_V} A_V' d A_V'  
= \int_0^{A_V} {d N \over d\eta} { d \eta \over d A_V} A_V' d A_V' \\ 
= \int_0^{A_V} {d N \over d\eta} {A_V' \over A_V'} d A_V' 
= \int_0^{A_V} {d N \over d\eta} d A_V' 
\end{eqnarray*} 
For any power-law range of the PDF with exponent s, we thus expect the 
integral to have the corresponding slope s+1.  The main difference 
between the cumulative mass functions therefore stems from the 
different slopes $s$ of the PDF tails. NGC3603 with the lowest 
exponent shows a shallow mass function providing a large fraction of 
high-mass contributions. However, the correspondence here with the 
slope of the power-law tail slope of the PDF is not as good, probably 
because we left out a larger fraction of high column 
densities.  Comparing this result with the models of Federrath \& 
Klessen (2013) that were used in Kainulainen et al.  (2013), it 
becomes obvious that NGC3603 fits very well into the scenario where 
the higher fraction of dense gas is due to compressive driving. In the 
extreme case of simulation with only compressive modes, the flat 
slopes are even nearly independent of the star-formation efficiency 
(SFE). In the case of mixed driving, magnetic fields have a small impact 
but more importantly, a higher SFE flattens the slope.  Without going 
into a detailed comparison to these models, which will be done in a 
forthcoming paper, we conclude that external compression plays the 
most important role in shaping the column density structure of 
NGC3603, consistent with what we deduced from the PDF and column density 
profile study. 
 
It is interesting to compare the behavior of Carina and Auriga, both 
having about the same slope for the power-law tail of the PDF but a 
different column density range of the power-law fit.  Consequently, 
the cumulative mass functions of both maps have the same slope over 
the common mass range, but the low-mass star-forming regions breaks 
off at lower masses/column densities. In a linear mass scaling, the 
break-off behavior, i.e. contributions around the upper column 
density where the power-law tail ends, appear more prominent. They 
represent the few densest cores that might be the places of 
high-mass or low-mass star formation. 
  
\section{Conclusions}   
 
In this paper, we present column density maps of four molecular clouds 
obtained from {\sl Herschel}. These are the Auriga and Maddalena 
clouds, forming low-mass stars, and Carina and NGC3603 that are 
UV-illuminated, high-mass star-forming GMCs.  We present a simple 
method of correcting for line-of-sight (LOS) contamination from fore- or 
background clouds by removing a constant layer of emission and then study 
the effect of this procedure on the resulting column density 
structure and the probability distribution functions of column 
density in observations and simulations. Our findings are: 
 
\noindent $\bullet$ The PDFs for all observed clouds become broader,  
the peak shifts to lower column densities, and the power-law tail of  
the PDF for higher column densities flattens after correction. \\ 
\noindent $\bullet$ We simulated the effect of LOS contamination by  
generating a PDF with typical observational parameters, consisting  
of a lognormal part and a power-law tail, and then `contaminated'  
this PDF by adding a constant level to all map values.  The simulations  
show that LOS-contamination strongly compresses the  
lognormal part of the PDF, consistent with what is observed for  
distant clouds. The peak of the PDF and the value where the PDF  
turns from lognormal into a power-law tail (\av(DP)) increases, and 
the slope of the power-law tail becomes steeper.\\ 
\noindent $\bullet$ We created plots in which for various contamination  
levels the change in the PDF width ($\sigma_\eta$) and slope of the  
power-law tail ($s$) can be assessed. For a contamination of \dav=2 mag,  
$\sigma_\eta$ is already reduced by more than a factor two, and $s$  
has steepened from $-2.0$ to $-2.4$. \\ 
\noindent $\bullet$ The convolution of the map with the beam has only 
a minor effect on the high-column density tail of the PDF and no 
influence on the lower density lognormal distribution. Resolution 
effects are thus less important when analyzing column density 
PDFs and cumulative mass functions. \\ 
\noindent $\bullet$ All observed PDFs that were corrected for  
LOS-contamination have a lognormal part for low column densities  
with a peak at \av\, $\sim$2 mag and a deviation point (DP) from the  
lognormal at \av(DP)$\sim$4--5 mag. For higher column densities, all  
PDFs have a power-law tail with an average slope of --2.6$\pm$0.5. \\ 
\noindent $\bullet$ Assuming an equivalent spherical density distribution  
$\rho \propto r^{-\alpha}$, this average slope corresponds to an 
exponent $\alpha_{PDF}$ = 1.9$\pm$0.3 consistent with the view that the  
gas in the power-law tail is dominated by self-gravity (local  
free-fall of individual cores and global collapse of gas on larger  
scales, such as filaments). \\ 
\noindent $\bullet$ Our PDF study suggests that there is a common 
column density break at \av\ $\sim$4--5 mag for all cloud types where 
the transition between supersonic turbulence and self-gravity takes 
place. Our value is lower than the one found by Froebrich \& Rowles 
(2010) with \av\ = 6.0$\pm$1.5 mag but consistent with cloud simulations  
of self-gravitating turbulent gas (Ward et al. 2014). \\ 
\noindent $\bullet$ The average column density for low-mass SF 
regions (1.8 10$^{21}$ cm$^{-2}$) is slightly lower than the one for 
high-mass SF clouds (3.0 10$^{21}$ cm$^{-2}$). Because of the small 
sample (four clouds), and the uncertainties introduced by cropping 
effects, it is not clear whether this difference is 
statistically significant.  \\ 
\noindent $\bullet$ Radial column density profiles of three clouds  
in our sample show a distribution that is compatible with (global)  
gravitational collapse for Auriga and Carina (slopes correspond to  
$\alpha$ = 1.5 and 1.8), but requires an additional compression  
process for NGC3603 ($\alpha$ = 2.5). We suggest that compression  
from the expanding ionization fronts from the associated \hii\ region  
leads to a forced collapse. This is consistent with the higher mass fraction at higher  
column densities in the cumulative mass function and corresponds well with  
numerical simulations based on compressive driving. \\ 
\noindent $\bullet$ In view of the differences observed for the slopes 
of the power-law tail of the PDF and the variation in the run of the 
column density profiles, we find that the column density structure of 
clouds forming low-mass stars and the ones forming 
massive stars are not the same.  \\ 
\noindent $\bullet$ The cumulative mass distributions for high-mass 
SF regions is shallower than the one for low-mass SF clouds for high 
column densities, indicating a higher concentration of dense gas in 
smaller cloud volumes. 
  
\begin{acknowledgements}    
  N.S., S.B., and P.A. acknowledge support by the ANR-11-BS56-010 
  project ``STARFICH''. \\  
  V.O., N.S., P.G., and  R.S.K. acknowledge subsidies from the 
  Deutsche Forschungsgemeinschaft, priority program 1573 (``Physics of 
  the Interstellar Medium''). \\ 
  R.S.K.  acknowledges support by the collaborative research project SFB 
  881 (``The Milky Way System'', subprojects B1, B2, and B5), and 
  support from the European Research Council under the European 
  Communities Seventh Framework Program (FP7/2007-2013) via the ERC 
  Advanced Grant STARLIGHT (project number 339177). \\ 
  C.F.~acknowledges funding provided by the Australian Research Council's Discovery Projects  
  (grants~DP130102078 and~DP150104329). \\  
  T.Cs.  acknowledges financial support from the ERC Advanced Grant GLOSTAR under contract 
  no. 247078. \\ 
  Ph. Andr\'e acknowledges financial support from the ERC Advanced Grant ORISTARS under contract 
  no. 291 294. \\ 
  V.O. and N.S. acknowledge support by the Deutsche 
  Forschungsgemeinschaft, DFG, through project number 0S 177/2-1. \\ 
  We thank an anonoymous referee for the detailed comments that 
  improved the paper. 
\end{acknowledgements}

 
\appendix 
\renewcommand{\thesection}{A.\arabic{section}} 
\section{Effects of binning and cropping on the PDF}   \label{app-a}    
\setcounter{figure}{0} \renewcommand{\thefigure}{A.\arabic{figure}}  
 
Figure~\ref{auriga-bin} shows the effect on the PDF of varying the 
binsize.  We started first from the original column density map on a 
14$''$ grid, which approximately corresponds to Nyquist sampling for 
an angular resolution of the data of 36$''$.  It becomes obvious that 
there is basically no effect on the derived values of \av(peak), 
\av(DP), and slope. For the fit, we fixed only the width of the PDF in 
order to avoid the lower column density peak, arising from a 
seperate component, being taken into account. A binsize of 0.1 turns out 
to be the best choice for our data because finer sampling (0.05) 
increases the statistical noise for the higher column density range, 
and a lower sampling (0.2) smoothes out features in the PDF that can 
still be significant. 
  
\begin{figure*}[!htpb] \label{auriga-bin}     
\centering    
\includegraphics [width=7.5cm, angle={90}]{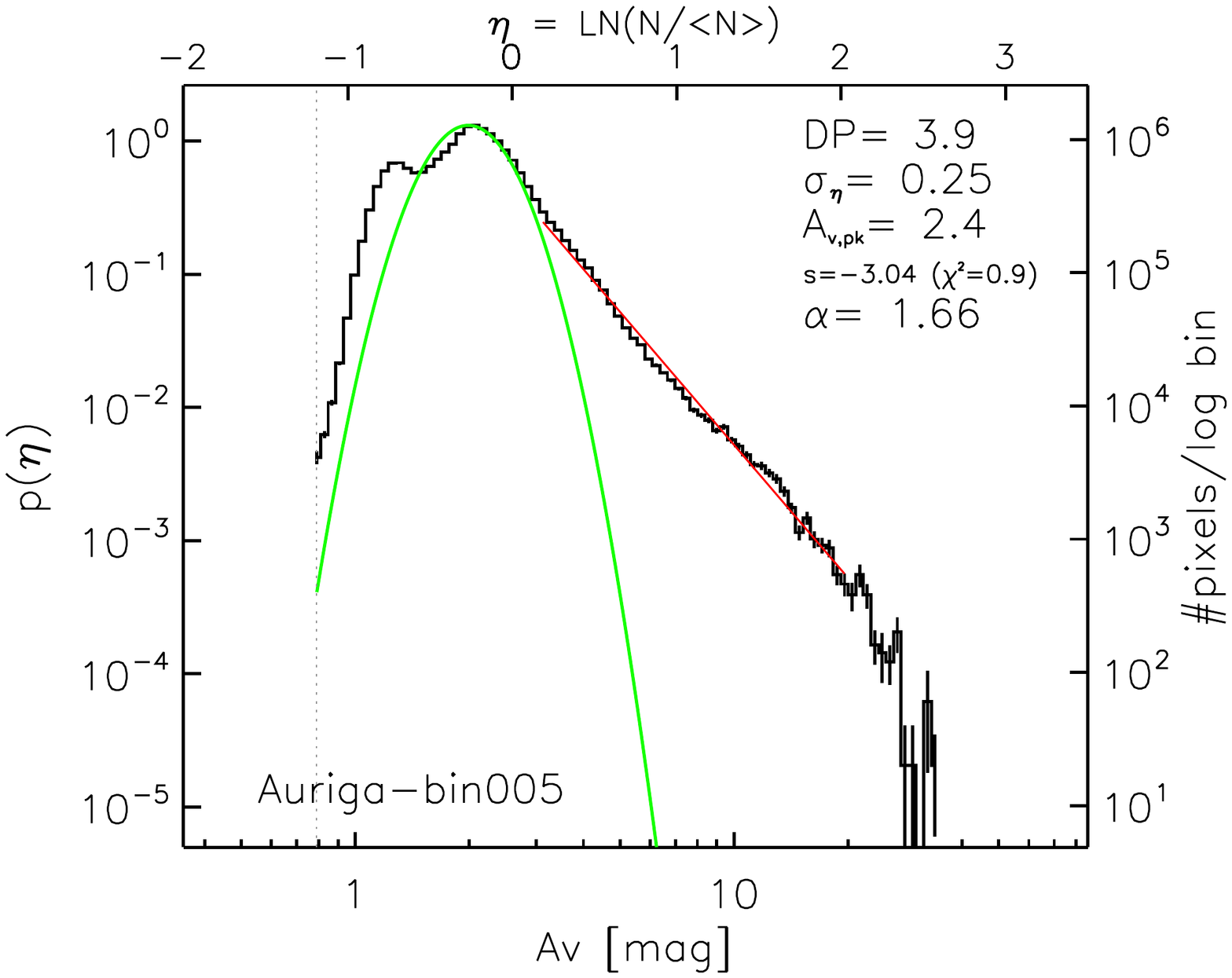}    
\hspace{-1.5cm} \includegraphics [width=7.5cm, angle={90}]{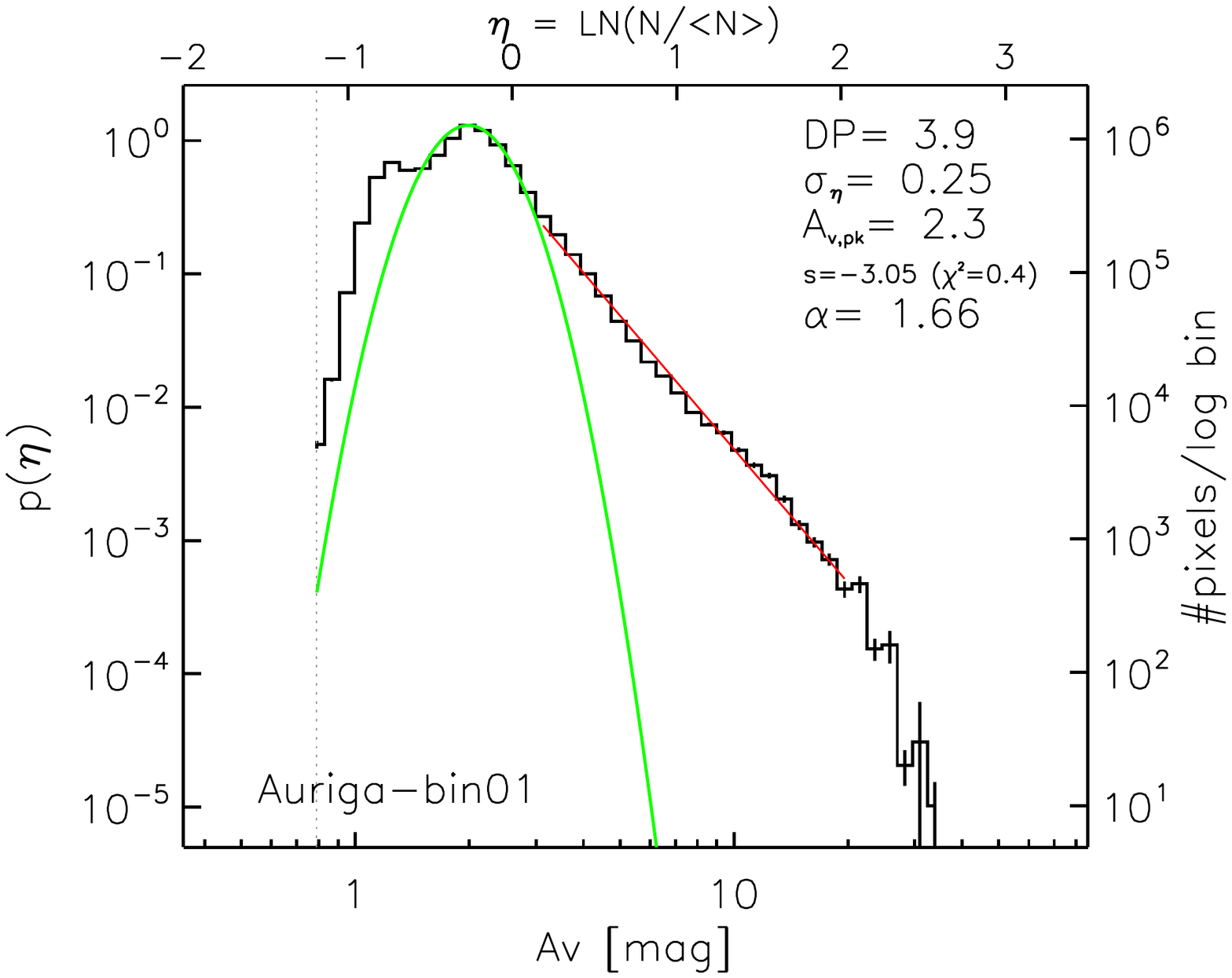}  
 
\vspace{-2.0cm}   
\includegraphics [width=7.5cm, angle={90}]{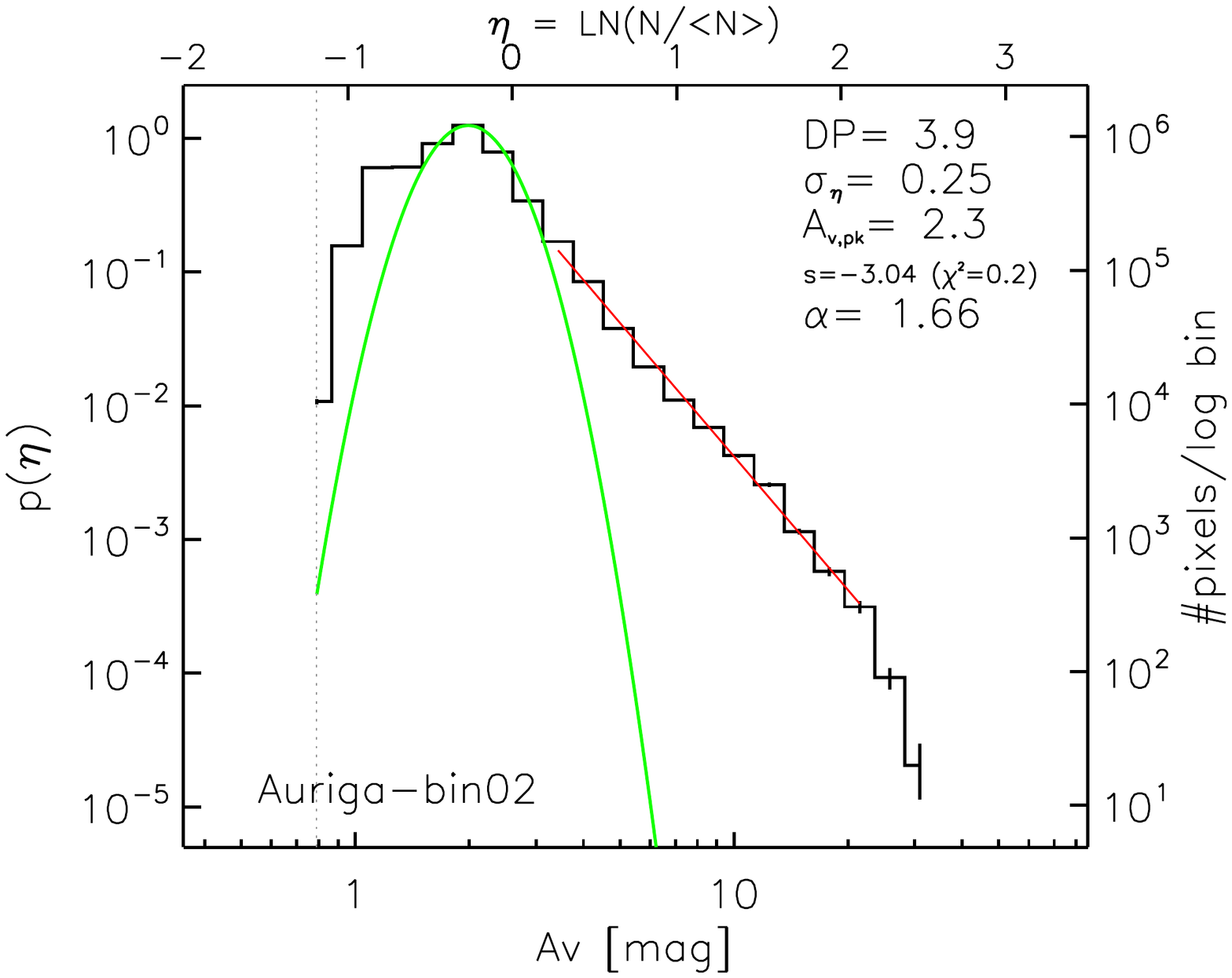}    
\hspace{-1.5cm} \includegraphics [width=7.5cm, angle={90}]{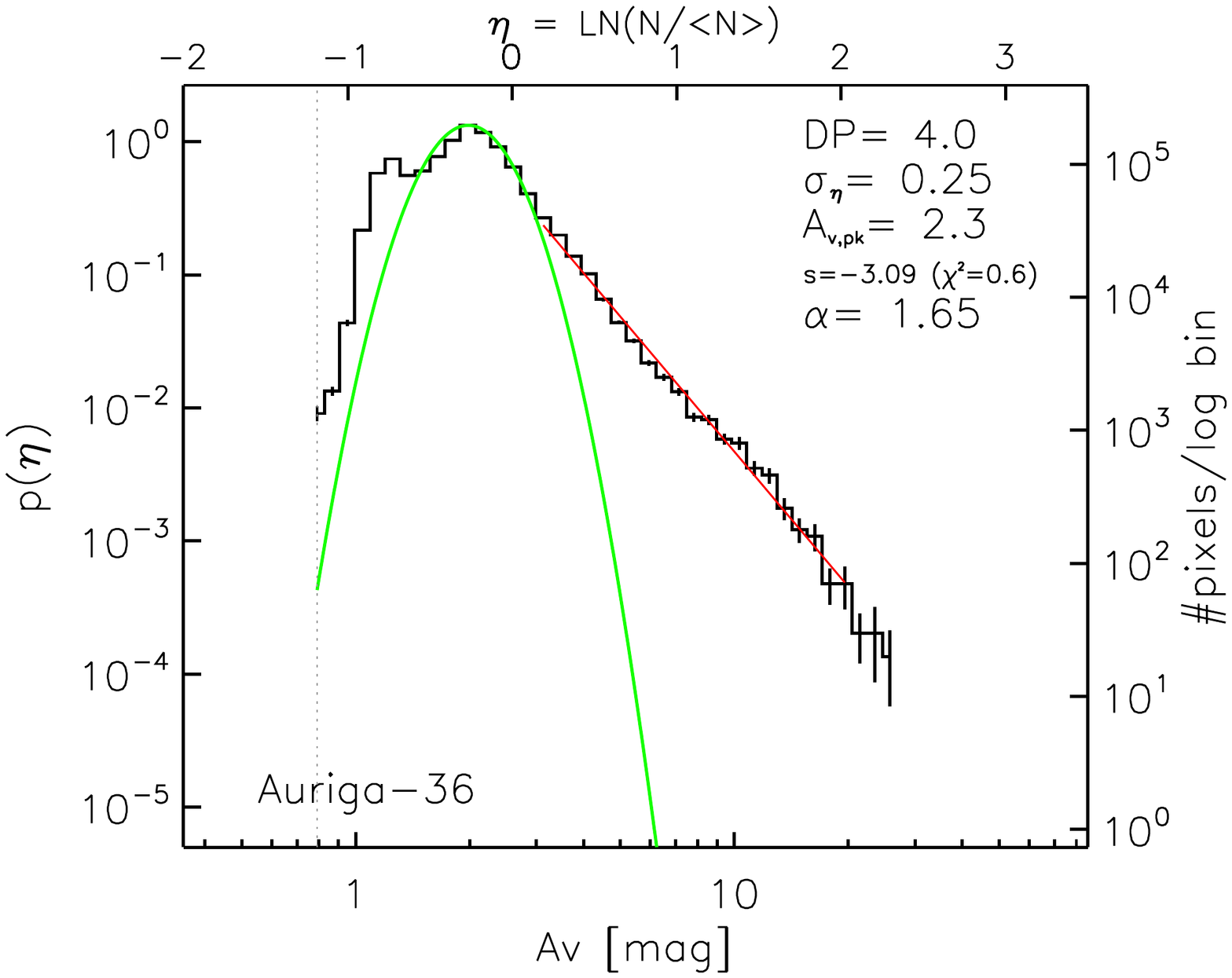}    
\caption[] {PDFs obtained from the orginal (not LOS-corrected) Auriga 
  column density map with different bin sizes (0.05, 0.1, and 0.2) and 
  a PDF with a binsize of 0.1 but on a grid of 36$''$ (lower, right 
  panel).  The vertical dashed line indicates the noise level of the 
  map. The left y-axis gives the normalized probability $p(\eta)$, the 
  right y-axis the number of pixels per log bin. The upper x-axis is 
  the visual extinction and the lower x-axis the logarithm of the 
  normalized column density.  The green curve indicates the fitted PDF 
  (we fixed the width of the PDF in order to avoid the low column 
  density component being included).  The red line indicates a power-law 
  (linear regression) fit to the high \av\, tail (the start- and 
  end-point were fixed to the same values for each PDF).  Inside each 
  panel, we give the value where the PDF peaks (A$_{pk}$), the 
  deviation point from lognormal to power-law tail (DP), the 
  dispersion of the fitted PDF ($\sigma_\eta$), the slope $s$ and the 
  $X^2$ of the fit, and the exponent $\alpha$ of an equivalent 
  spherical density distribution.} 
\end{figure*}    
 
\begin{figure*}[!htpb]   \label{auriga-crop}     
\centering    
\includegraphics [width=15cm, angle={0}]{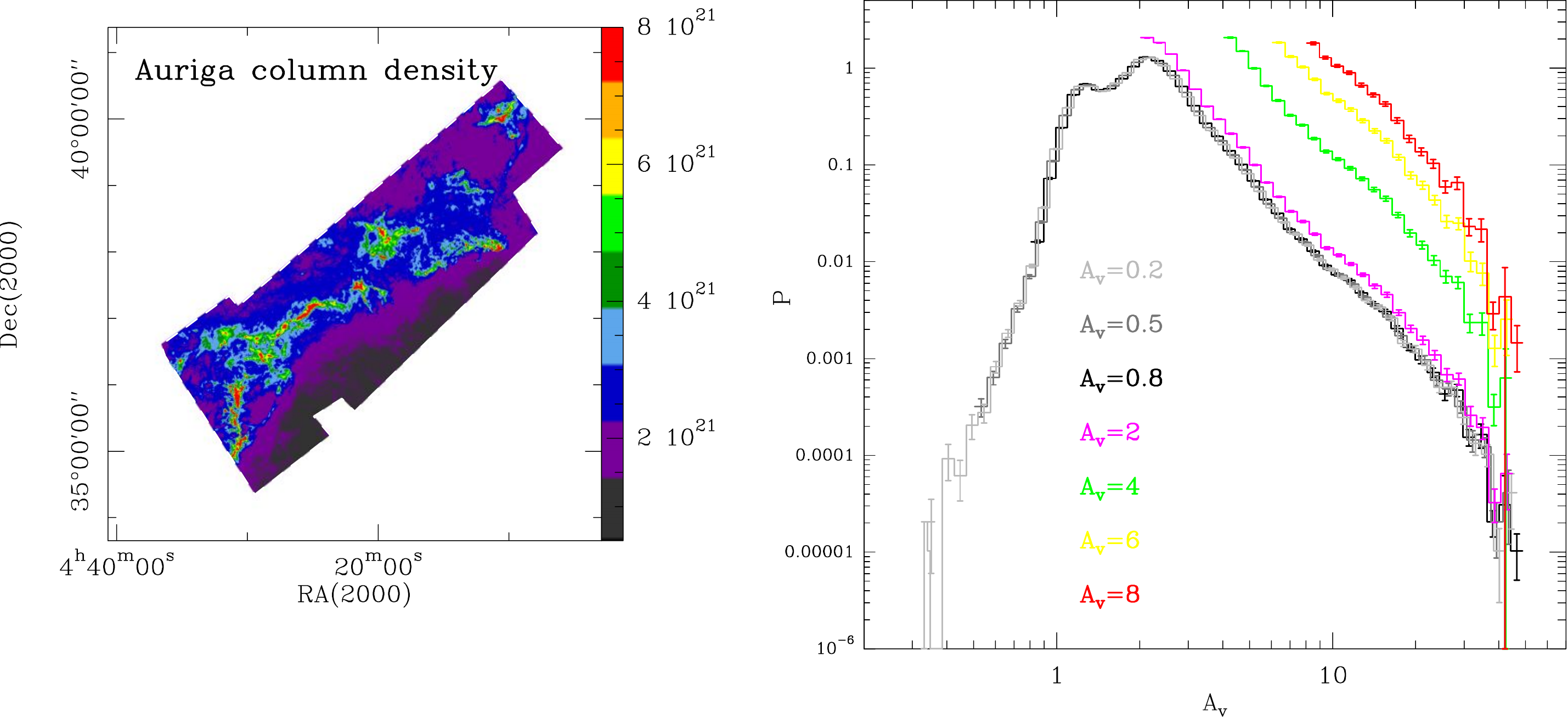}    
\caption[] {Left: the column density map 
    of Auriga (without correction for LOS-contamination), showing the 
    emission distribution at various thresholds of \av. The colors 
    correspond to the PDFs (right panel) constructed {\sl above} 
    \av\, levels of 0.8, 2, 4, 6, and 8.  The y-axis gives the 
    normalized probability $p(\eta)$, and the x-axis the visual 
    extinction.} 
\end{figure*}    
 
In Fig.~A.2 we investigate the effect of `field-cropping', which is how 
the PDF shape changes when only pixels above a certain threshold are 
considered, using the Auriga cloud as an example. First, we observe 
that for high column density ranges, i.e. above the peak of the 
lognormal distribution, the PDF is only represented by a power-law 
tail all with the same slope, independent of \av\,-threshold. 
Small variations in the PDF shape are due to the normalization process. 
(The average column density changes for each threshold-value.) As a 
consequence, PDFs obtained by `cropping' images at high column 
densities, such as PDFs for IRDCs in IR-quiet molecular clouds, 
should thus be strongly dominated by the power-law tail. This was 
recently confirmed observationally by Schneider et al. (2014). 
Toward the low column density range, we constructed PDFs above 
threshold values of \av=0.2 (noise level), \av=0.5, and \av=0.8 
(background level). We do not observe a significant effect on the PDF 
shape when we change the threshold, though it is clear that for the 
lowest column densities, the clouds in our {\sl Herschel} maps are not 
completely sampled. 
 
 
\renewcommand{\thesection}{B.\arabic{section}} 
\section{Herschel maps of molecular clouds } \label{app-b}    
 
\setcounter{figure}{0} \renewcommand{\thefigure}{B.\arabic{figure}}  
 
All maps presented in this paper -- and in the furthcoming ones -- 
were treated in the same way with regard to data reduction and 
determination of the column density maps. For the data reduction of 
the {\sl Herschel} wavelengths, we used the HIPE10 pipeline, including 
the destriper task for SPIRE (250, 350, 500 $\mu$m), and HIPE10 and 
scanamorphos v12 (Roussel et al.  \cite{roussel2013}) for PACS (70$''$ 
and 160$''$, Poglitsch et al.  \cite{poglitsch2010}).  The SPIRE 
(Griffin et al.  \cite{griffin2010}) maps include the turnaround-data 
(when the satellite changed mapping direction for the scan) and were 
calibrated for extended emission. 
 
The procedure for making column density and temperature maps is 
independent of the data reduction process and follows the scheme 
described in Schneider et al. (\cite{schneider2012}).  The column 
density maps were determined from a pixel-to-pixel graybody fit to the 
red wavelength of PACS (160 $\mu$m) and SPIRE (250, 350, 500 $\mu$m). 
We did not include the 70 $\mu$m data because we focus on the cold 
dust and not UV-heated warm dust (see section 4.1 in Russeil et al. 
(\cite{russeil2013}) for more details on this issue).  The maps were 
first convolved to have the same angular resolution of 36$''$ and 
then regridded on a 14$''$ raster. All maps have an absolute flux 
calibration, using the {\sc zeroPointCorrection} task in HIPE10 for 
SPIRE and IRAS maps for PACS.  
 
The correction for SPIRE works in such a way that that a
cross-calibration with Planck maps at 500 and 350 $\mu$m is performed.
We emphasize that such a correction is indispensable for accurately
determining column density maps.  For the SED fit, we fixed the
specific dust opacity per unit mass (dust+gas) approximated by the
power law $\kappa_\nu \, = \,0.1 \, (\nu/1000 {\rm GHz})^\beta$
cm$^{2}$/g and $\beta$=2 and left the dust temperature and column
density as free parameters. We checked the SED fit of each pixel and
determined from the fitted surface density the H$_2$ column density.
For the transformation H$_2$ column density into visual extinction, we
used the conversion formula N(H$_2$)/\av = 0.94$\times$10$^{21}$
cm$^{-2}$ mag$^{-1}$ (Bohlin et al.  1978).  The angular resolution of
the column density maps is $\sim$36$''$, and they are shown in
Figs.~B.1 to B.4, together with SPIRE 250 $\mu$m images.  We estimate
the final uncertainties in the {\sl Herschel} column density maps to
be around $\sim$30--40\%, mainly arising from the uncertainty in the
assumed form of the opacity law, and possible temperature gradients
along the line of sight (see Roy et al.  \cite{roy2014} for details on
the dust opacity law and Russeil et al.  \cite{russeil2013} for a
quantitative discussion on the various error sources).
  
\begin{figure*}[!htpb]  \label{fig-auriga}   
\centering    
\includegraphics [width=8cm, angle={0}]{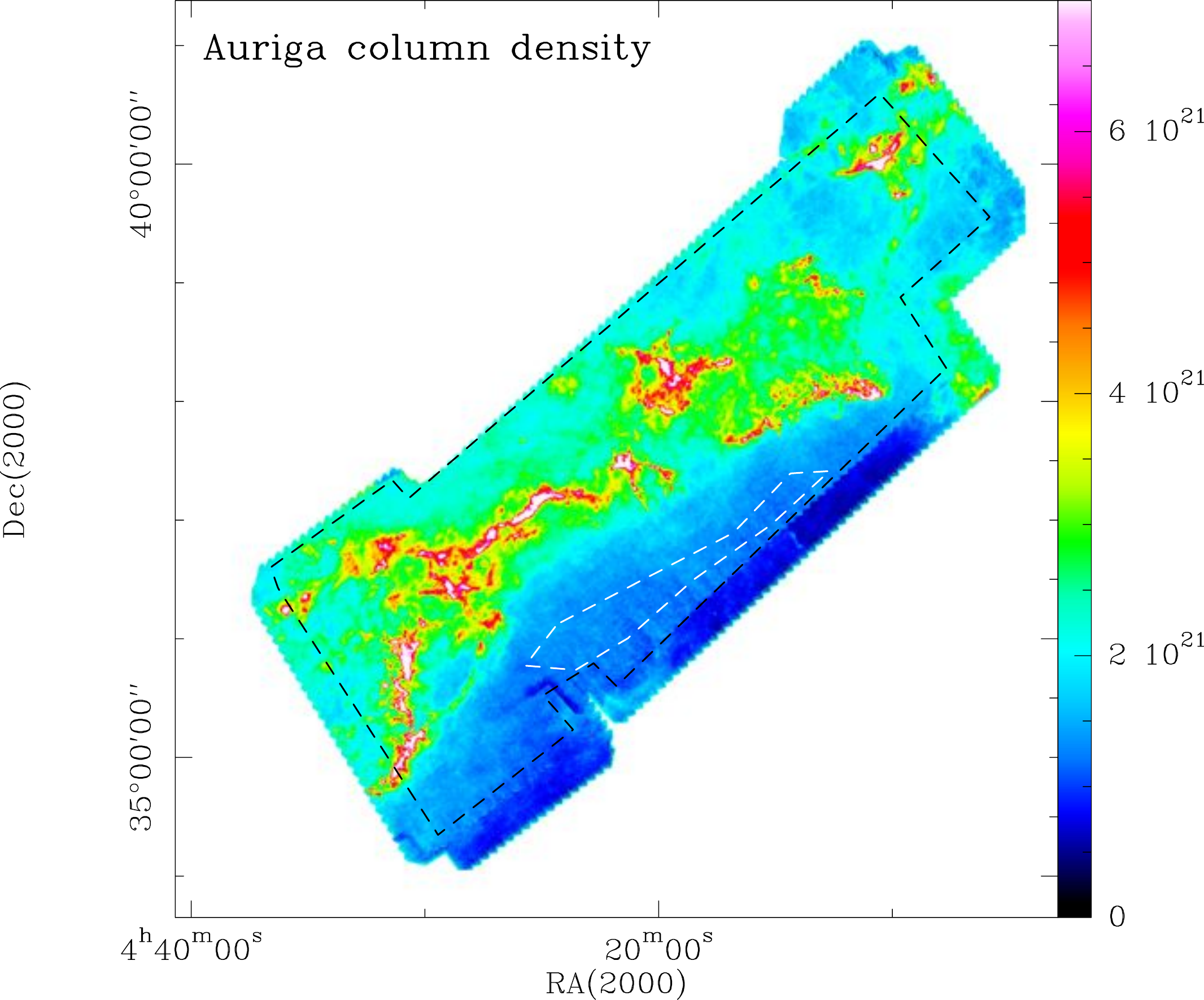}    
\hspace{1cm} 
\includegraphics [width=6.9cm, angle={0}]{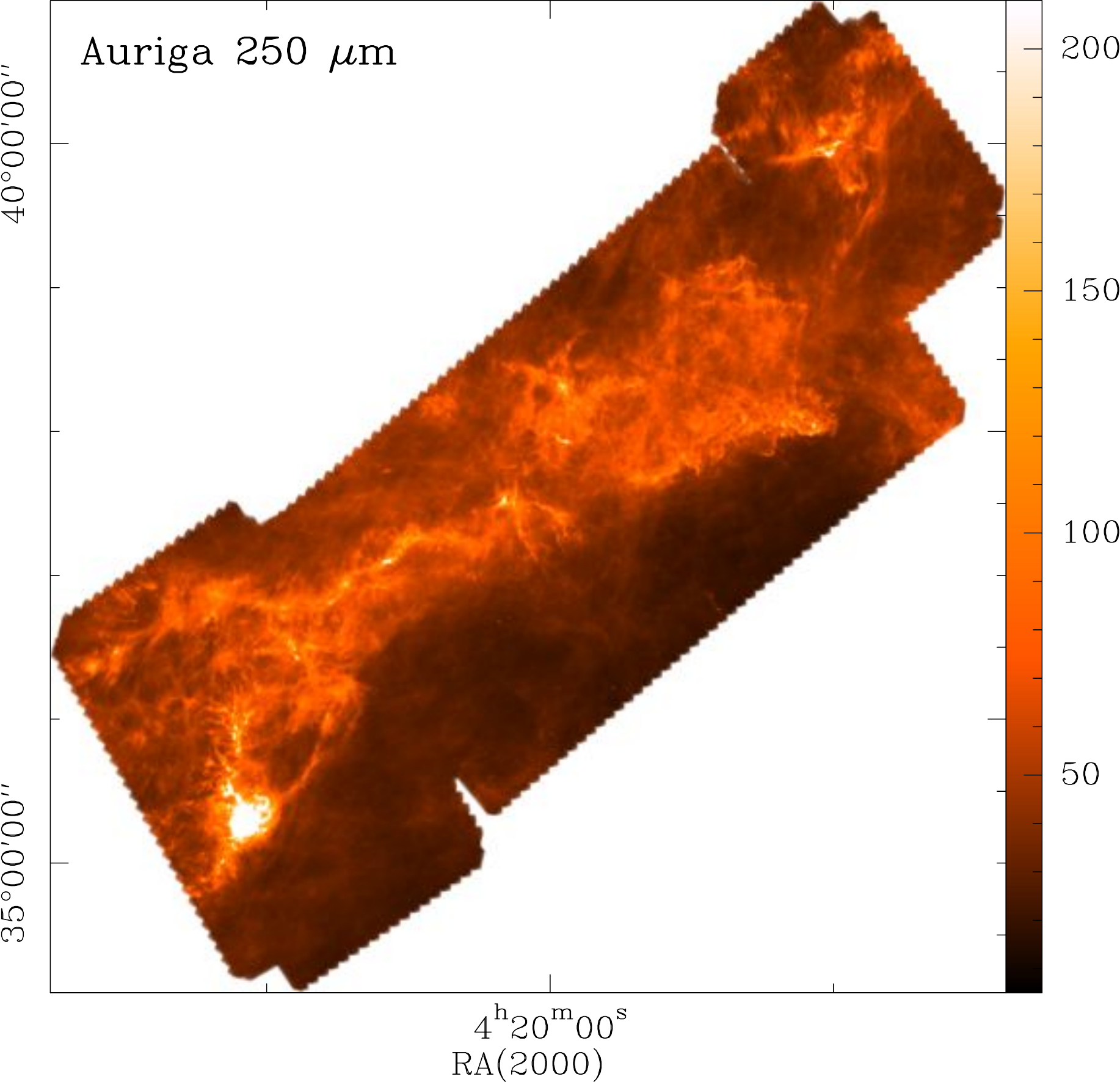}    
\caption[] {{\bf Left:} {\sl Herschel} column density map of the 
  Auriga cloud in [cm$^{-2}$].  The black dashed polygon indicates the 
  pixels from which the PDF was determined. This is the common overlap 
  region of SPIRE and PACS in which the column density map was 
  determined from the SED fit using the 4 wavelengths.  Outside this 
  polygon, the fit relies only on SPIRE and it less reliable. The 
  white polygon outlines the area used for determining the 
  background/foreground level of contamination of the map.  {\bf 
    Right:} SPIRE 250 $\mu$m map in units [MJy/sr]. } 
\end{figure*}    
 
\begin{figure*}[!htpb]  \label{fig-maddalena}   
\centering    
\includegraphics [width=8.5cm, angle={0}]{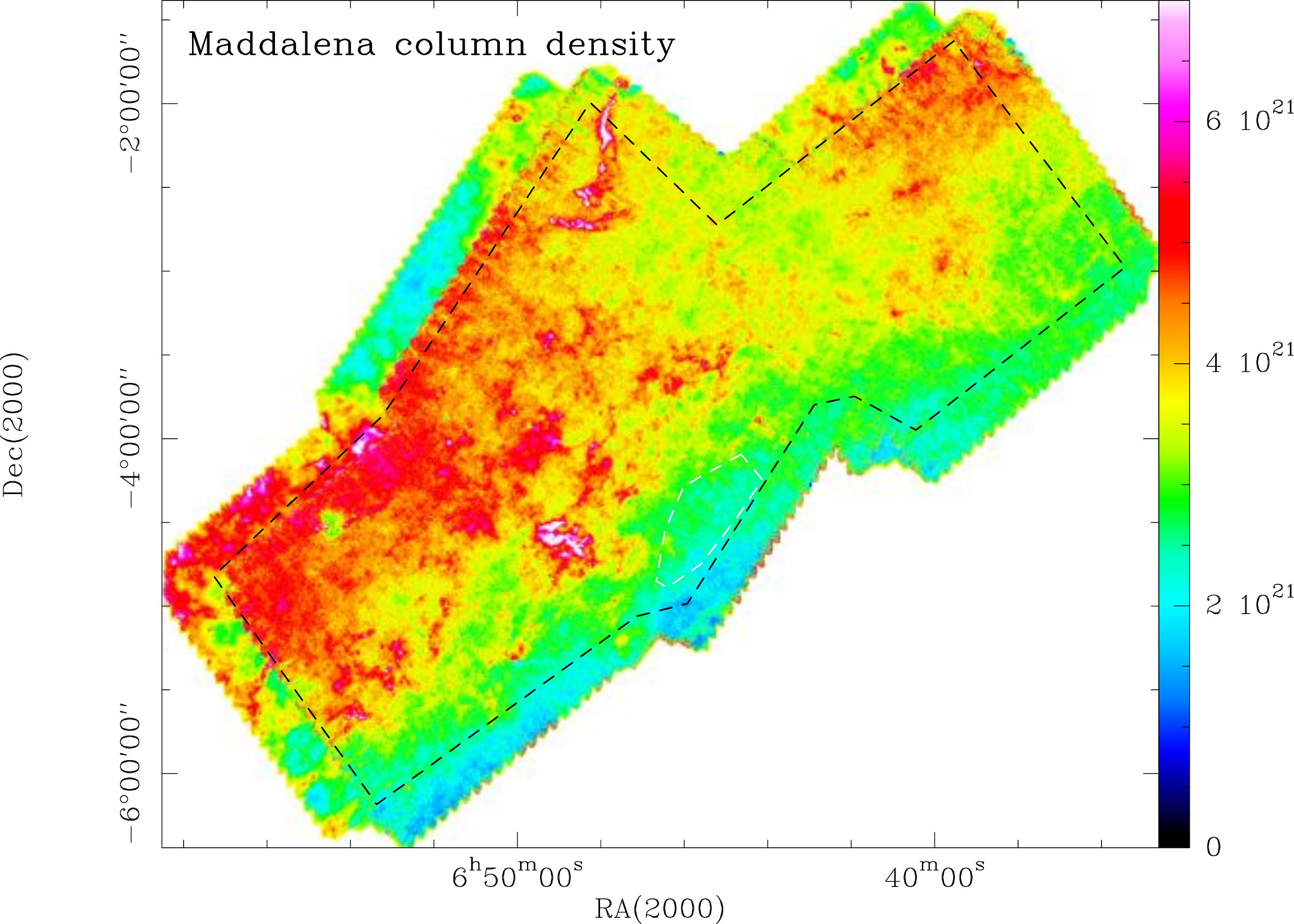}    
\hspace{1cm} 
\includegraphics [width=7.4cm, angle={0}]{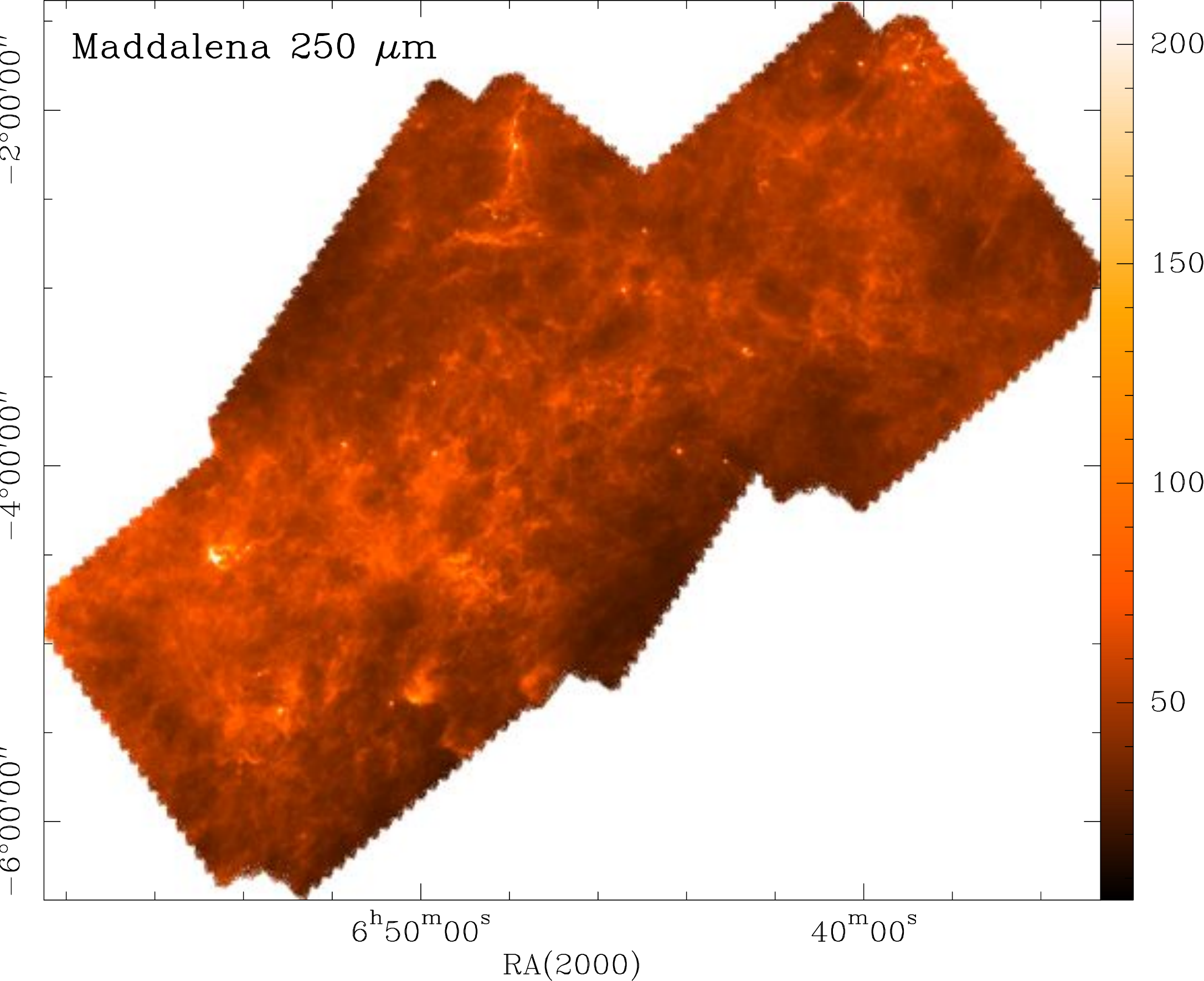}    
\caption[] {{\bf Left:} {\sl Herschel} column density map of the  
  Maddalena cloud in [cm$^{-2}$]. All other parameters as in Fig.~B.1. 
 {\bf Right:} SPIRE 250 $\mu$m map in units  
  [MJy/sr]. }  
\end{figure*}    
  
\begin{figure*}[!htpb]  \label{fig-ngc3603}   
\centering    
\includegraphics [width=9cm, angle={0}]{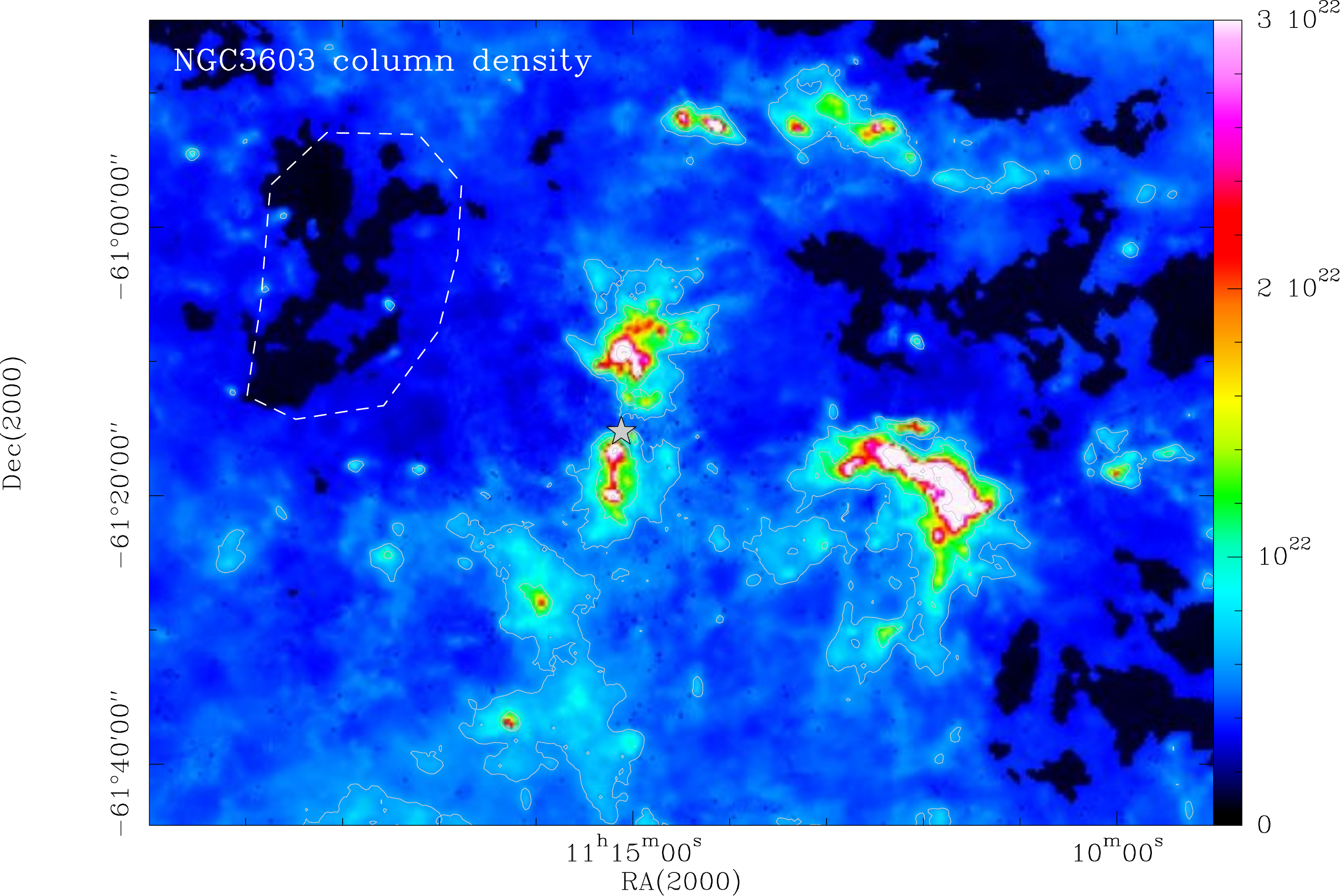}    
\includegraphics [width=9cm, angle={0}]{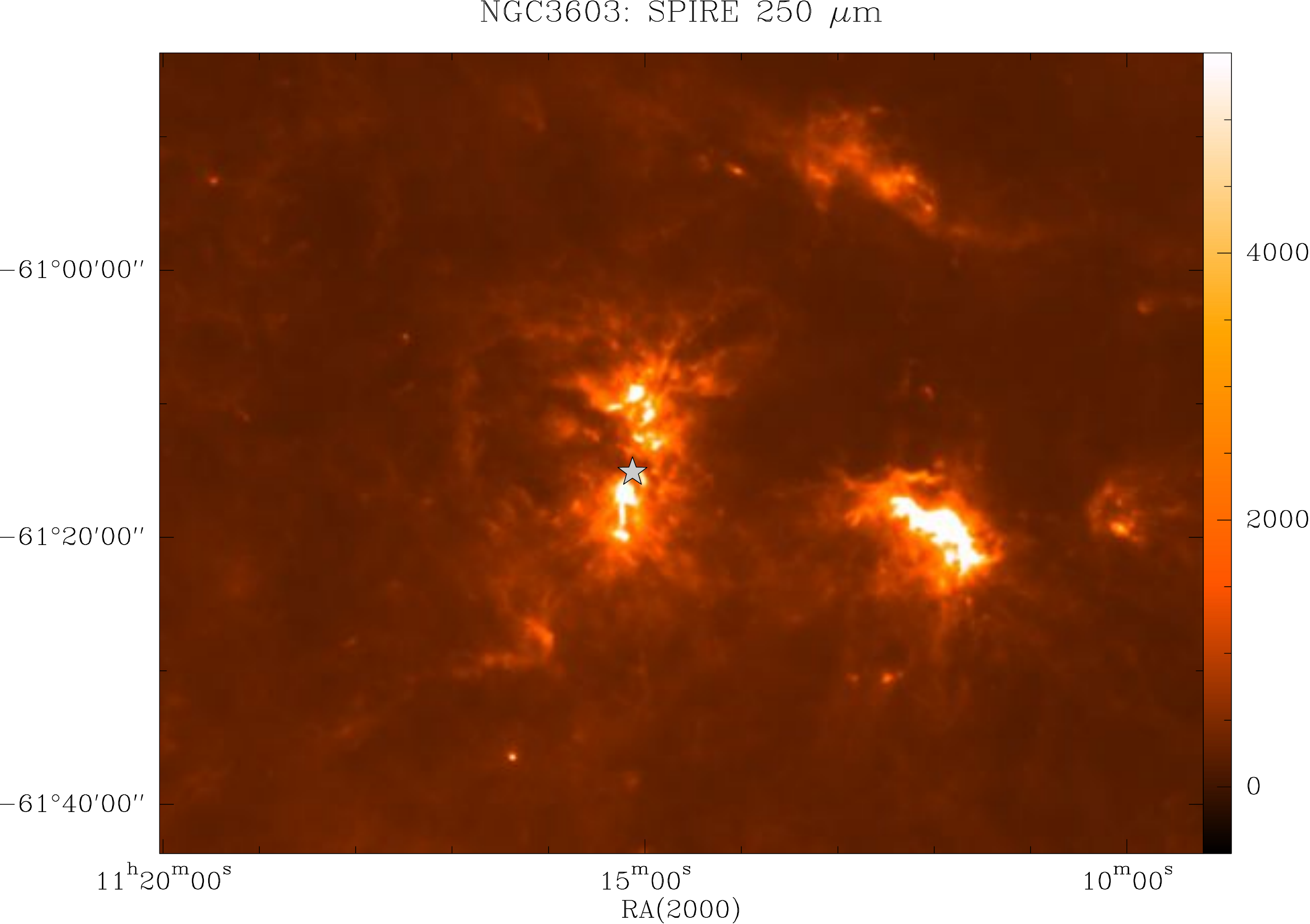}    
\caption[] {{\bf Left:} {\sl Herschel} column density map of NGC3603  
  in [cm$^{-2}$]. The contour levels are 3, 6, 10, and 20 10$^{21}$ cm$^{-2}$.  
All other parameters as in Fig.~B.1. {\bf Right:} SPIRE 250 $\mu$m map in units [MJy/sr].  
  The gray star indicates the location of the central cluster.}  
\end{figure*}    
   
\begin{figure*}[!htpb]  \label{fig-carina}   
\centering    
\includegraphics [width=8cm, angle={0}]{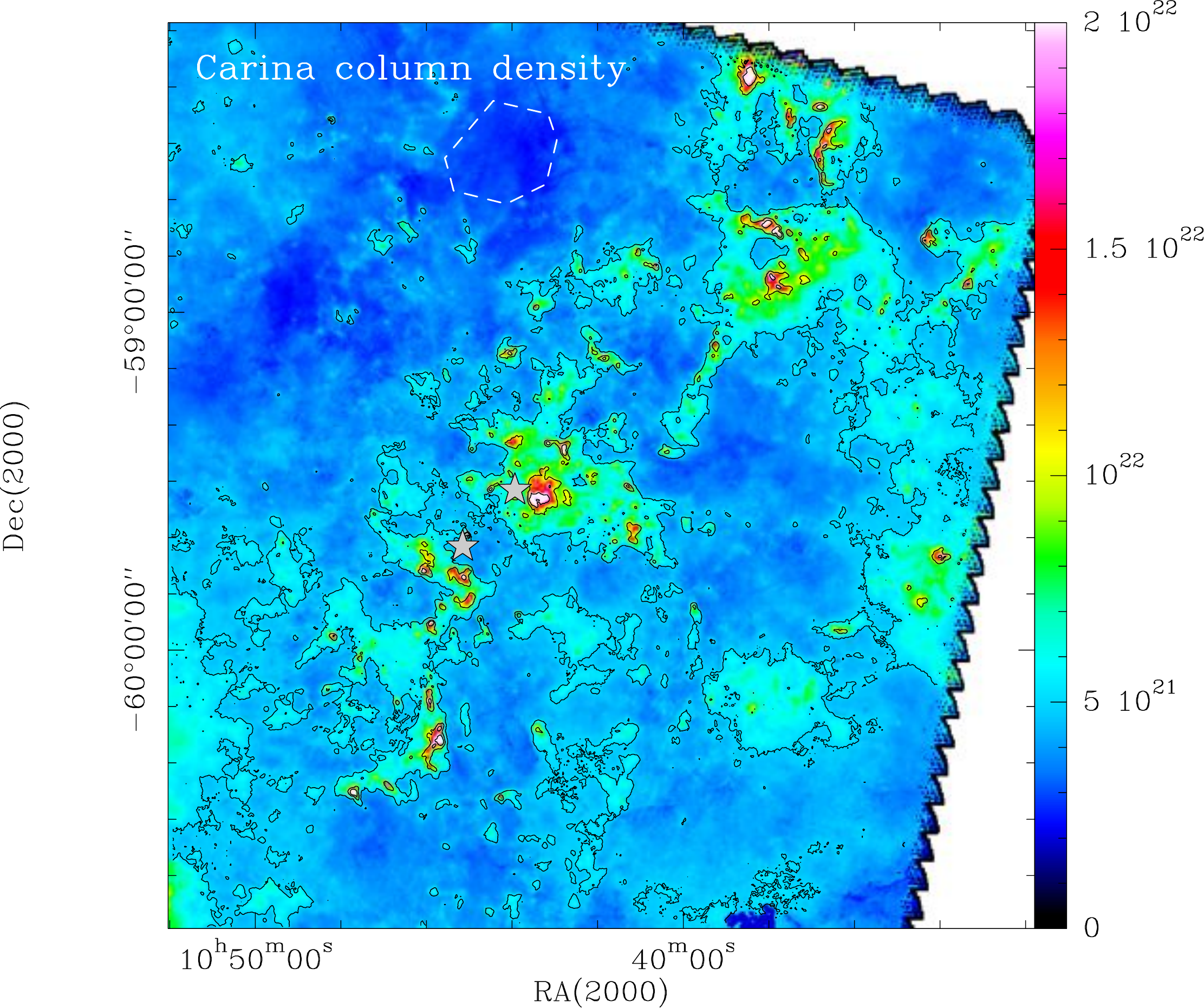}    
\hspace{1.5cm}   
\includegraphics [width=7.2cm, angle={0}]{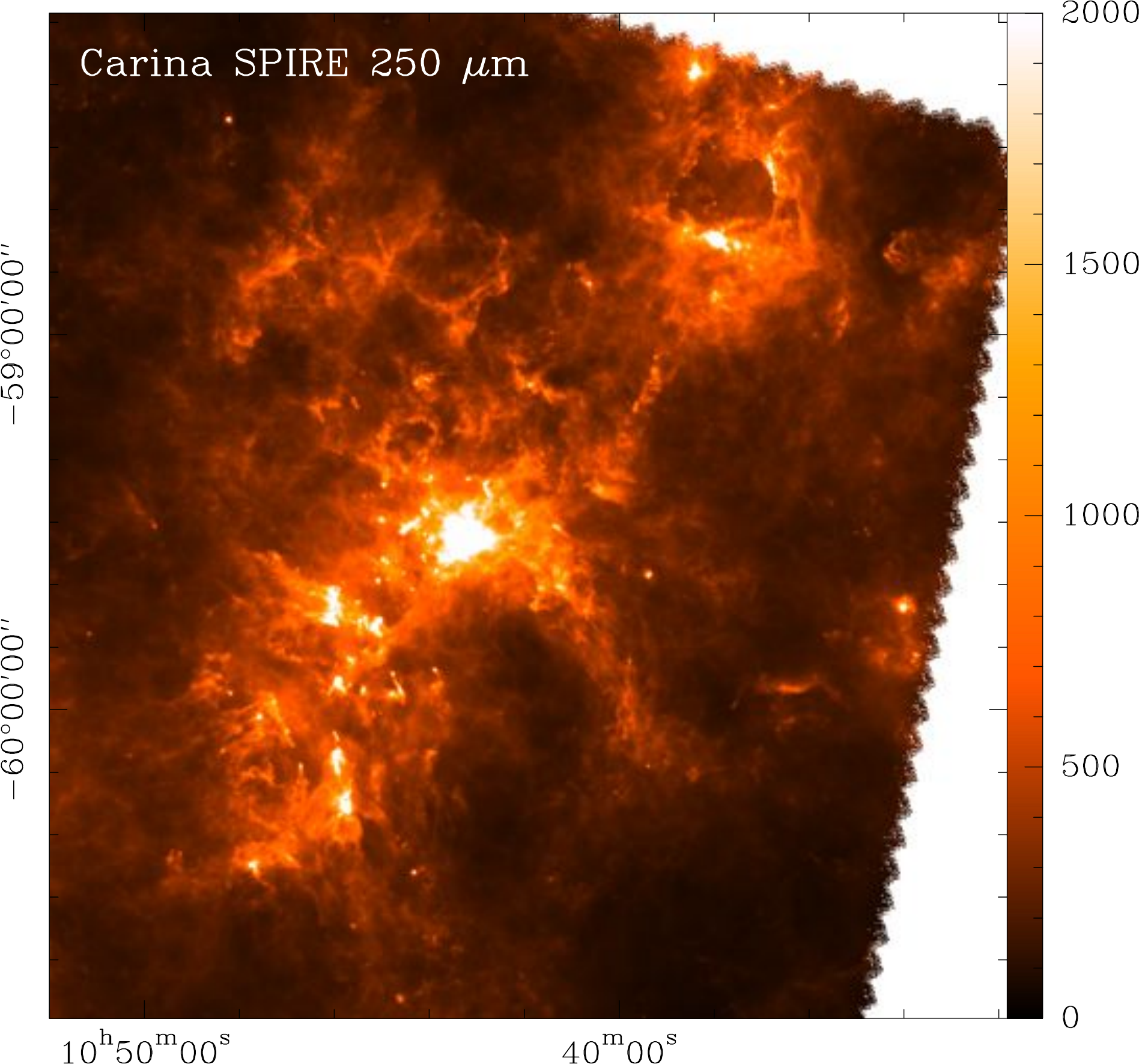}    
\caption[] {{\bf Left:} {\sl Herschel} column density map of Carina in  
  [cm$^{-2}$]. The contour levels are 5, 10, and 20 10$^{21}$ cm$^{-2}$.  
All other parameters as in Fig.~B.1. {\bf Right:} SPIRE 250 $\mu$m map in units [MJy/sr]. The  
  gray stars indicate the location of the OB clusters Tr14 and Tr16.}  
\end{figure*}

    

\end{document}